 \documentclass[preprint,3p,times,twocolumn]{elsarticle}
\usepackage[T1]{fontenc}

\usepackage{epsfig}
\usepackage{amsmath,amssymb,bm}
\usepackage{graphics}
\usepackage{epsfig}
\usepackage{graphicx}
\usepackage{verbatim}
\usepackage{xcolor}

\newcommand{\be}{\begin{equation}}
\newcommand{\ee}{\end{equation}}
\newcommand{\bea}{\begin{eqnarray}}
\newcommand{\eea}{\end{eqnarray}}
\newcommand{\avg}[1]{\langle{#1}\rangle}

\newcommand{\eq}[1]{(\ref{#1})}

\journal{Chaos, Solitons \& Fractals}

\begin{document}

\begin{frontmatter}



\title{Statistics of remote regions of networks}


\author[label1]{J. G. Oliveira}

\affiliation[label1]{organization={Departamento de F\'{\i}sica da Universidade de Aveiro $\&$ I3N},
            addressline={Campus Universit\'ario de Santiago}, 
            city={Aveiro},
            postcode={3810-193}, 
            country={Portugal}}

\author[label1,label2]{S. N. Dorogovtsev}

\affiliation[label2]{organization={Ioffe Physico-Technical Institute},
            city={St. Petersburg},
            postcode={194021}, 
            country={Russia}}

\author[label1]{J. F. F. Mendes}

\begin{abstract}

We delve into the statistical properties of regions within complex networks that are distant from vertices with high centralities, such as hubs or highly connected clusters. These remote regions play a pivotal role in shaping the asymptotic behaviours of various spreading processes and the features of associated spectra. 
We investigate the probability distribution $P_{\geq m}(s)$ of the number $s$ of vertices located at distance $m$ or beyond from a randomly chosen vertex in an undirected network. 
Earlier, this distribution and its large $m$ asymptotics $1/s^2$ were obtained theoretically for undirected uncorrelated networks \cite{dorogovtsev2003metric}. 
Employing numerical simulations and analysing empirical data, we explore a wide range of real undirected networks and their models, including trees and loopy networks, and reveal that the inverse square law is valid even for networks with strong correlations. 
We observe this law in the networks demonstrating the small-world effect and containing vertices with degree $1$ (so-called leaves or dead ends). 
We find the specific classes of networks for which this law is not valid. Such networks include the finite-dimensional networks and the networks embedded in finite-dimensional spaces. 
We notice that long chains of nodes in networks reduce the range of $m$ for which the inverse square law can be spotted. 
Interestingly, we detect such long chains in the remote regions of the undirected projection of a large Web domain.  

\end{abstract}




\end{frontmatter}




\section{Introduction}
\label{s1}

The particular focus within the research field of the statistical physics of complex networks is on the exploration and comprehension of the central regions of a network that house vertices with high centralities, such as hubs or highly connected clusters. A large number of various centrality measures are used for discovery and indexing of this important central part \cite{perra2008spectral,katz1953new,brin1998anatomy,page1999pagerank,freeman1977set,newman2005measure,newman2010networks,martin2014localization,newman2004finding,estrada2005subgraph}. 
The statistical properties of the remote regions of networks, distant from vertices with high centralities, are far less studied despite their significant role in the asymptotic behaviour of various spreading processes, random walks, and the features of associated spectra \cite{levin2017markov,samukhin2008laplacian,vukadinovic2002spectrum}.  
One of the simplest statistical characteristics of the remote regions of networks  is the shape of the tail of the distribution of shortest-path lengths \cite{dorogovtsev2022nature,krapivsky2001organization,dorogovtsev2002evolution,dorogovtsev2003metric,fronczak2004average,van2005distances,van2007distances,dorogovtsev2006degree,morohosi2010measuring,katzav2015analytical,katzav2018distribution,tishby2022mean}. Notably, in the networks demonstrating the small-world effect, this distribution approaches the delta function shape as the network size tends to infinity \cite{dorogovtsev2008organization} in contrast to finite-dimensional networks (``large worlds''). Hence, the exploration of remote network regions suggests a focus on networks that are large yet finite in size. In this paper we consider another statistical characteristic of the remote network regions in the giant connected component, namely, the probability distribution $P_{\geq m}(s)$ of the number $s$ of vertices located at distance $m$ or beyond from a randomly chosen vertex. 
This distribution was obtained theoretically in Ref.~\cite{dorogovtsev2003metric} for the configuration model of undirected uncorrelated networks with an arbitrary degree distribution. It was shown that the large $s$ asymptotics of this distribution, for sufficiently large $m$, follows the inverse square law 
\be
P_{\geq m}(s) \propto s^{-2}
\label{10}
\ee
if an uncorrelated network contains leaves (dead end vertices), that is vertices of degree $1$, while for the uncorrelated networks with the lowest non-zero degree of vertices equal $3$, the asymptotics does not follow this law. 
In the intermediate case of the uncorrelated networks with the lowest non-zero degree of vertices equal $2$, this asymptotics was obtained theoretically, but the range of its validity turned out to be narrow in reasonably sized networks, and so it is difficult to observe. 

One should emphasize that uncorrelated networks are rather special in the sense that they account only complex degree distributions, devoiding of the various correlations and short cycles that are prevalent in the majority of real-world networks. Furthermore, these compact networks, despite their locally tree-like organization, contain cycles, and hence they cannot be proper trees. This is why the theoretical asymptotics, Eq.~\eq{10}, was obtained only for a narrow class of networks. 
In this work we reveal that this inverse square asymptotics is actually observed in diverse real-world and synthetic undirected networks including strongly correlated networks, trees and loopy networks, demonstrating the small-world effect. These nets belong to the class of networks that is much wider than the uncorrelated networks. On the other hand, we indicate a set of networks for which this law is not valid. In particular, this set includes the finite-dimensional networks and the networks embedded in finite-dimensional metric spaces. 

Each distribution, plotted in each figure of this paper, was measured for one network realization through numerical computation of the number $s$ of vertices located at distance $m$ or beyond from each (and every) vertex in that specific realization of the network.

The paper is structured as follows. In Section~\ref{s2} we generate a number of synthetic networks, including an Erd\H{o}s--R\'enyi random graph, a random uniform tree, growing trees and loopy networks with various degree distributions and correlations, and measure in these networks the distribution $P_{\geq m}(s)$ and its asymptotics. 
In Section~\ref{s3} we analyze the structure of the remote regions of a set of real-world networks, including social networks, the Internet and the WWW, power grids, and road networks. 
We classify the networks in which the inverse square law is observed and indicate the networks in which it is not valid. 
In Section~\ref{s4} we discuss our results.



\section{Inverse square law in synthetic networks}
\label{s2}

\begin{figure*}[t]
\begin{flushleft}
\includegraphics[angle=0,width=225pt]{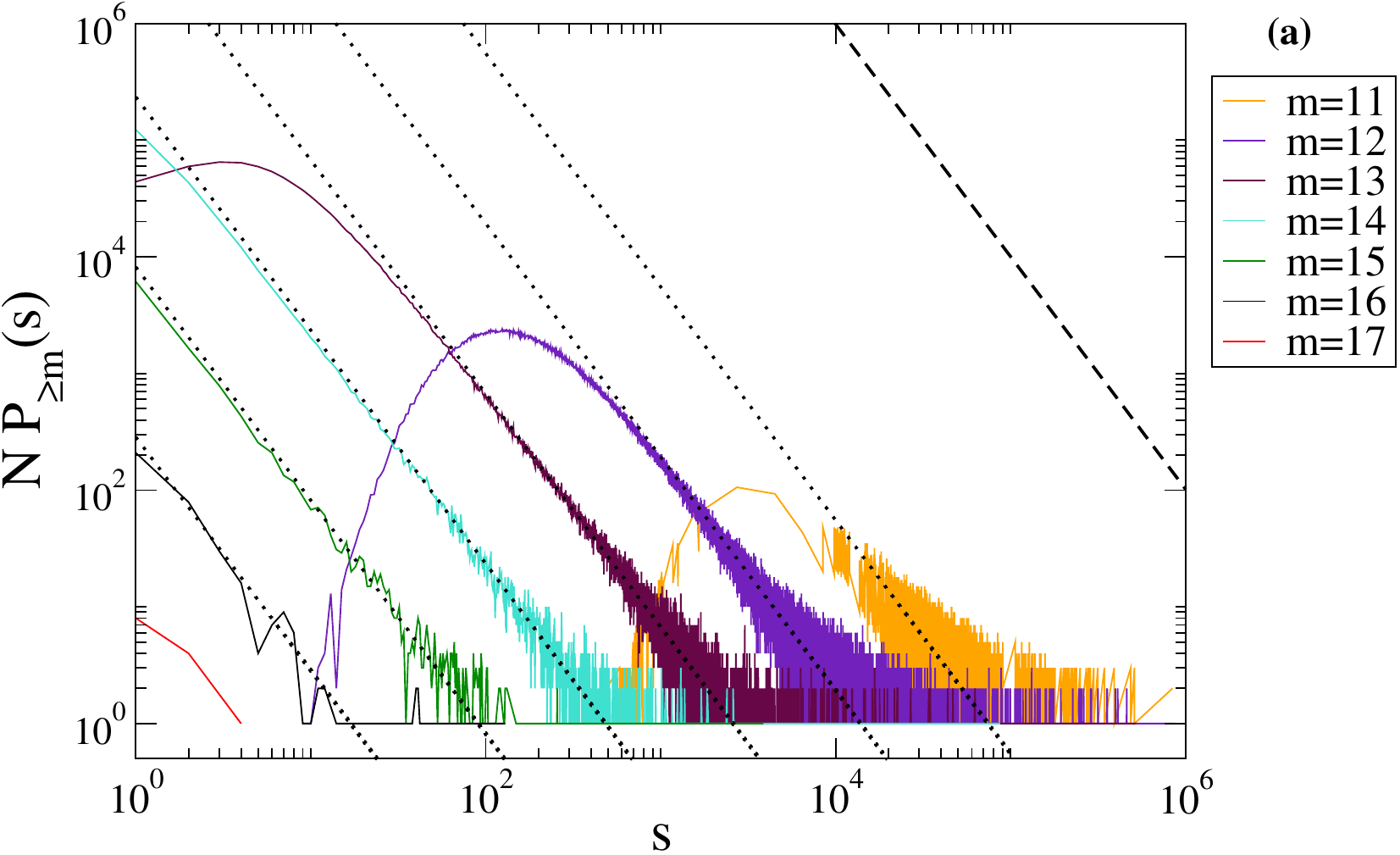}
\\[3pt]
\includegraphics[angle=0,width=225pt]{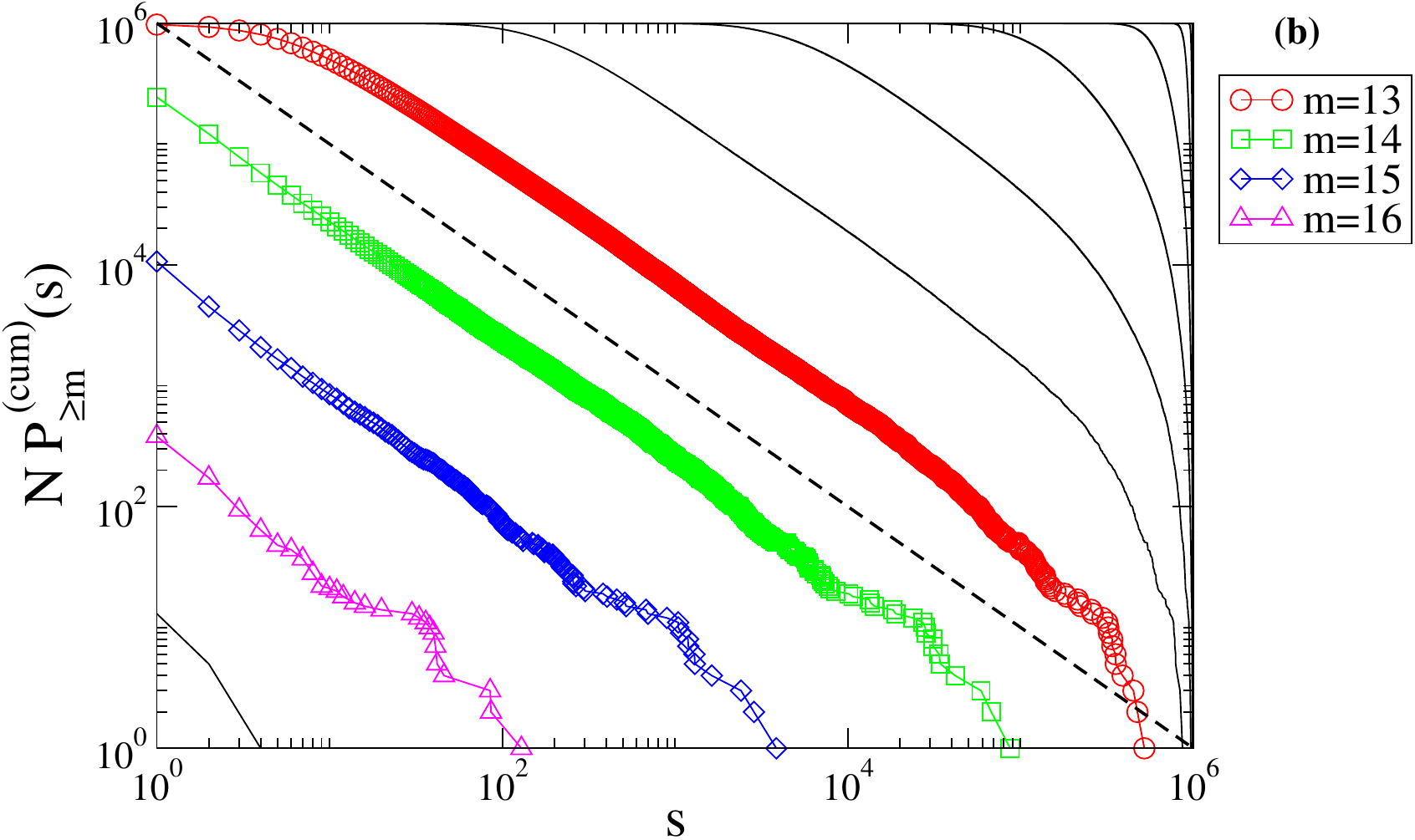}
\\[3pt]
\includegraphics[angle=0,width=225pt]{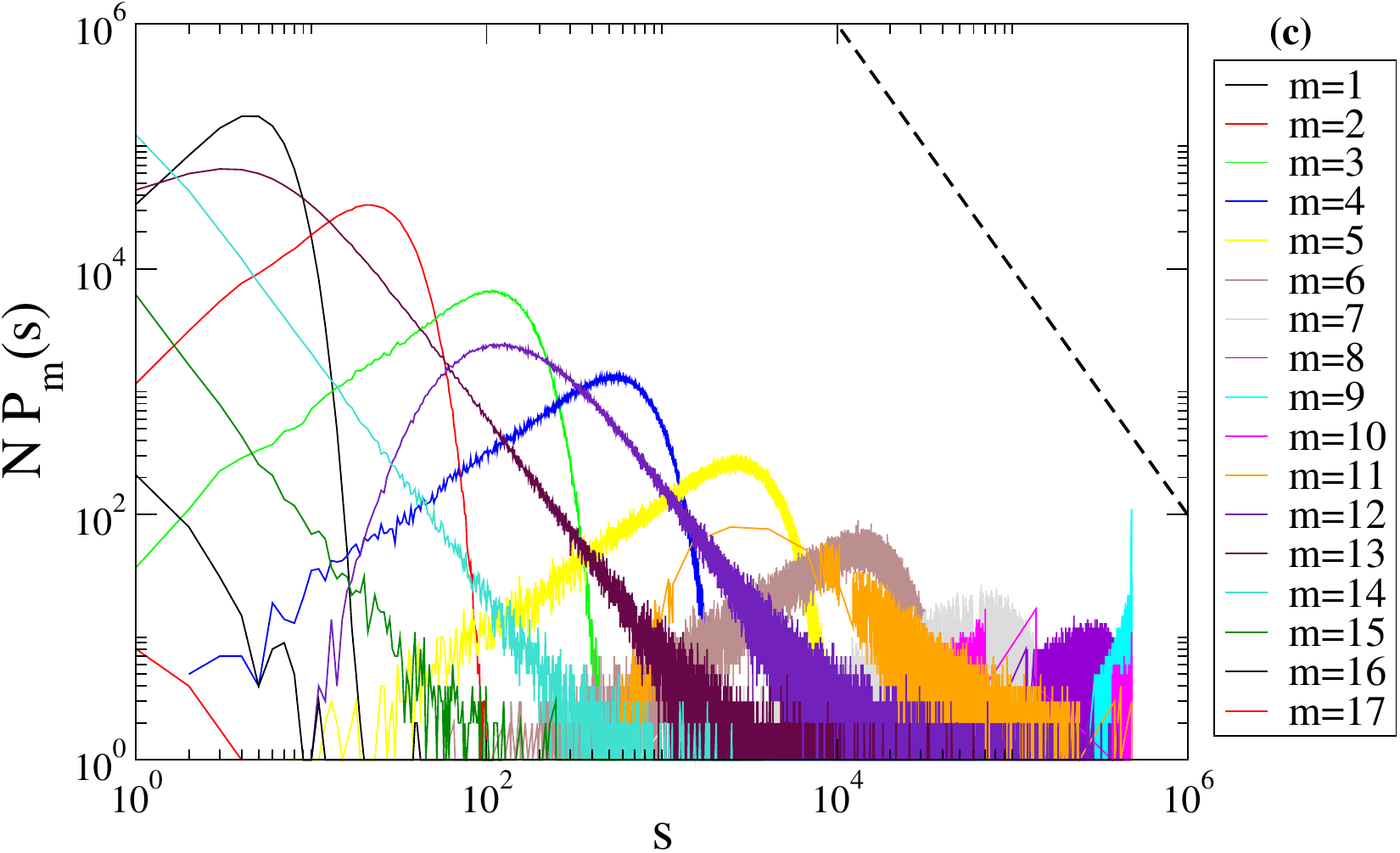}
\end{flushleft}
\caption{The statistics of the remote region of the Erd\H{o}s--R\'enyi random graph of $10^6$ vertices with the average vertex degree $\avg{q}=5$. (a) Distribution $P_{\geq m}(s)$ for different $m$.
The dotted lines show the theoretical asymptotics 
provided by Eq.~\protect\eq{20}. Dashed line has slope $-2$.
(b) Cumulative distribution $P^\text{(cum)}_{\geq m}(s)$ for different $m$. Dashed line has slope $-1$.
(c) Distribution $P_m(s)$ for different $m$. Dashed line has slope $-2$.
}
\label{f1}
\end{figure*}

It is natural to start our study with an Erd\H{o}s--R\'enyi random graph as the classical paradigm for random networks, being an uncorrelated network with a Poisson degree distribution. Figure~\ref{f1}(a) shows the distributions $P_{\geq m}(s)$ for different $m$ observed in the Erd\H{o}s--R\'enyi random graph of $10^6$ vertices, each pair of which is interconnected with a probability $p$, where the average degree of a vertex $\avg{q} \cong pN$ is $5$.  
For the sake of comparison, for each $m$ in the plot we indicate the corresponding theoretical asymptotics from Ref.~\cite{dorogovtsev2003metric}:  
\be
P_{\geq m}(s) \cong N \frac{z_c^{-1-n_0} B^2}{\Gamma(\alpha+1)}  z_c^m s^{-2}
. 
\label{20}
\ee
In this asymptotics, 
\be
n_0 = \frac{\ln [\avg{q}(\avg{q}-1)N]}{\ln \avg{q}}
\label{30}
\ee
and 
\bea
z_c &=& \avg{q} X_\infty < 1
, 
\label{40}
\\[3pt]
\alpha &=& -\frac{\ln (\avg{q} X_\infty)}{\ln \avg{q}}
, 
\label{50}
\eea
where 
$X_\infty = 1 - S_G$ ($S_G$ is the relative size of the giant connected component in the network) is the solution of the equation  
\be
X_\infty = e^{\avg{q}(X_\infty-1)}
. 
\label{60}
\ee
Finally, the $k\to\infty$ limit of the recursion 
\be
X_{k+1} = e^{\avg{q}(X_k -1)}
,
\label{70}
\ee
where the initial value is $X_0 = 1 - \delta$, $\delta\to0$, provides the constant $B$ in Eq.~\eq{20}, 
\be
B = \lim_{k\to\infty} (X_k-X_\infty)(\delta \avg{q}^k)^\alpha
.
\label{80}
\ee
The function $\Gamma(x)$ in Eq.~\eq{20} is the gamma function. 
Similar formulas describe the asymptotics of $P_{\geq m}(s)$ for the uncorrelated networks containing vertices of degree $1$. Notice an excellent agreement between the measured distribution and the theoretical one.  
It will be more convenient to observe the cumulative distribution $P^\text{(cum)}_{\geq m}(s) = \sum_{u \geq s} P_{\geq m}(u)$, for which this law, Eq.~\eq{10}, corresponds to the $1/s$ asymptotics, see Fig.~\ref{f1}(b). 
Furthermore, Fig.~\ref{f1}(c) shows the distribution $P_m(s)$ of the number $s$ of vertices located at distance $m$ from a randomly chosen vertex for different $m$. One can see that for sufficiently large $m$, the distribution $P_m(s)$ is close to $P_{\geq m}(s)$. 


\begin{figure*}[t]
\begin{flushleft}
\includegraphics[angle=0,width=225pt]{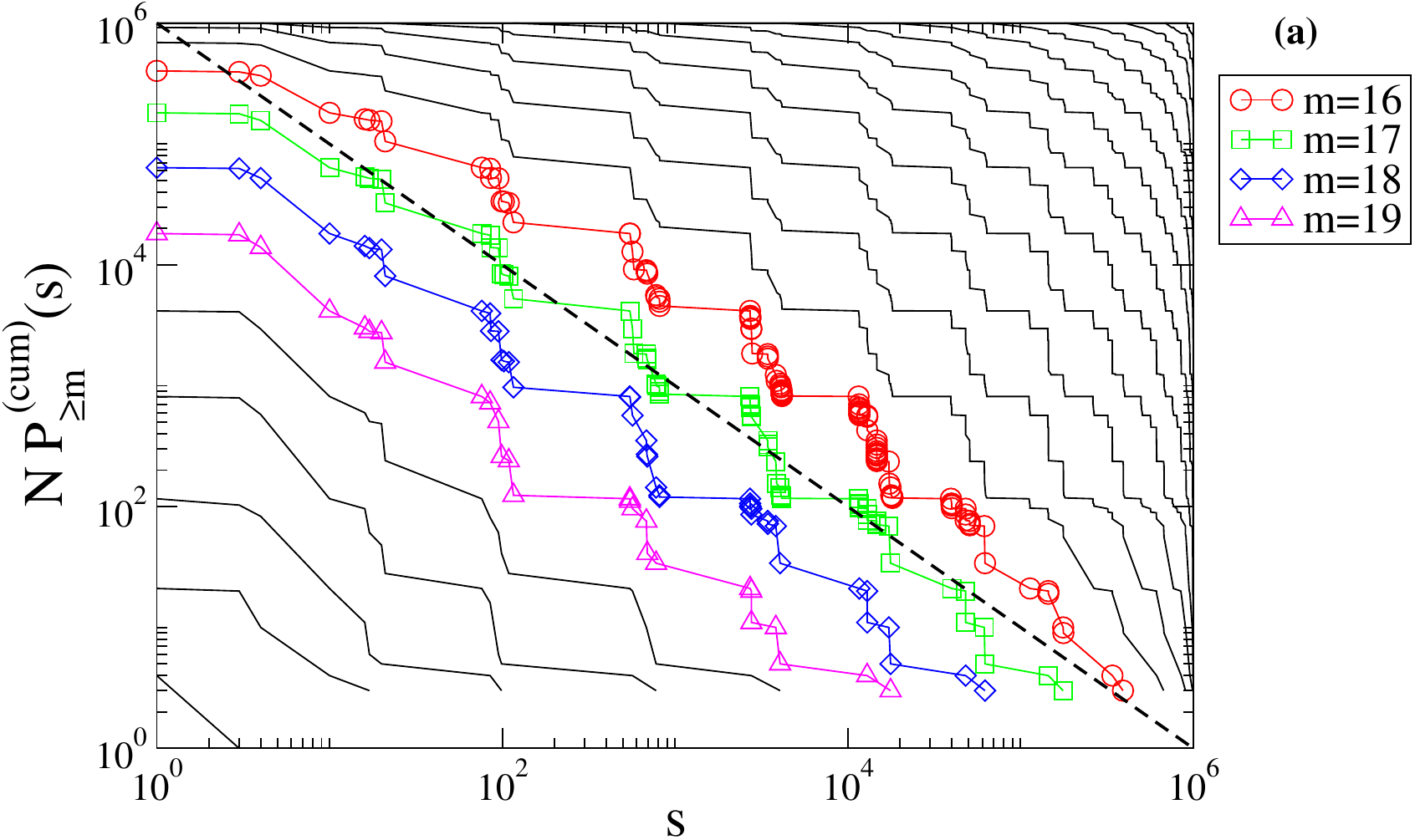}
\\[3pt]
\includegraphics[angle=0,width=225pt]{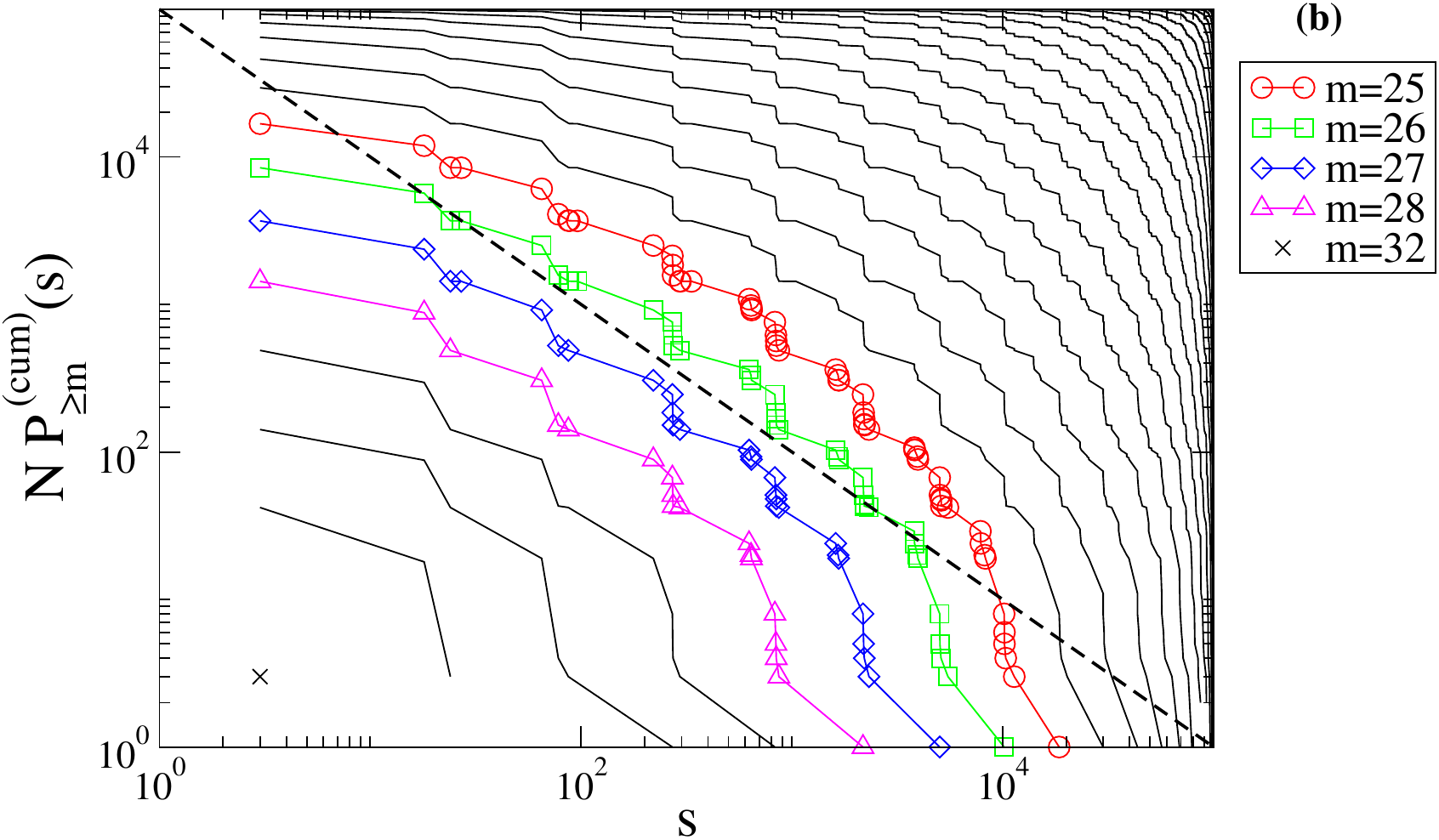}
\\[3pt]
\includegraphics[angle=0,width=225pt]{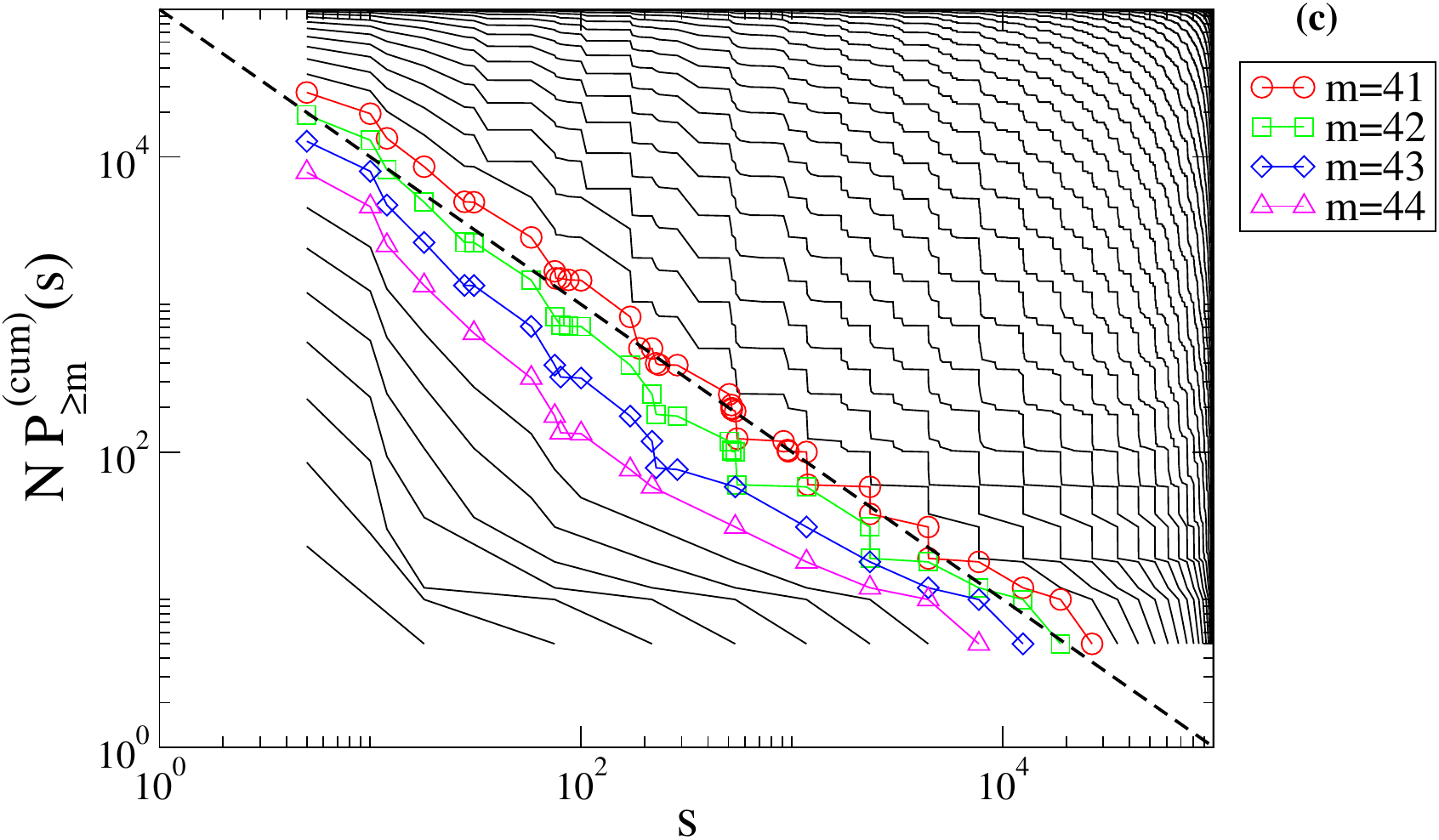}
\end{flushleft}
\caption{Cumulative distribution $P^\text{(cum)}_{\geq m}(s)$ of three random growing trees. Dashed lines have slope $-1$.
(a) A scale-free recursive tree of $10^6$ vertices whose growth is driven by the linear preferential attachment, $\text{Prob}(q_i) \propto q_i + A$, $A = -0.8$, where $q_i$ is the degree of vertex $i$. The degree distribution decays as $q^{-\gamma}$, $\gamma=3+A$. 
(b) A scale-free recursive tree of $10^5$ vertices generated by the Barab\'asi--Albert model (proportional preference). The degree distribution decays as $q^{-\gamma}$, $\gamma=3$. 
(c) A random recursive tree of $10^5$ vertices generated by progressive attachment of new vertices to randomly chosen vertices. Its degree distribution is exponential. 
}
\label{f2}
\end{figure*}

Let us now consider synthetic correlated networks. First we explore three recursive trees: the growth of two of them is driven by the linear preferential attachment algorithm, and hence they are scale-free, with the degree distribution exponents $\gamma=2.2$ and $3$ (Barab\'asi--Albert model---proportional preferential attachment), and the third is the random recursive tree, for which the degree distribution is exponential ($\gamma=\infty$). The first random tree has disassortative correlations between the degrees of the neighbouring vertices, the second has weak correlations, and the third has assortative correlations. All these growing random trees are small worlds. 
Figure~\ref{f2} demonstrates that the cumulative distributions $P^\text{(cum)}_{\geq m}(s)$ at sufficiently large $m$ decay as $1/s$.   


\begin{figure*}[t]
\begin{flushleft}
\includegraphics[angle=0,width=225pt]{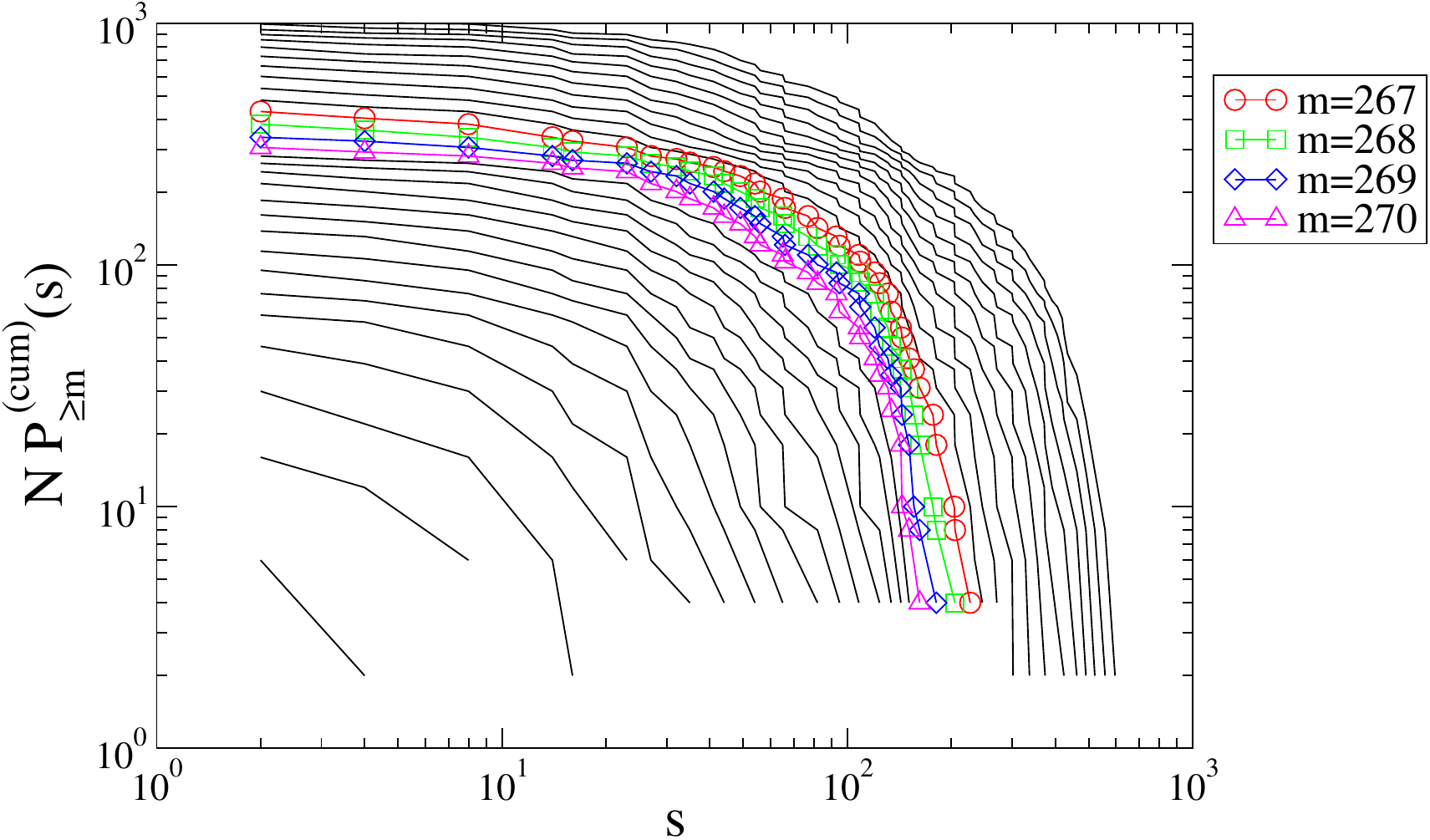}
\end{flushleft}
\caption{Cumulative distribution $P^\text{(cum)}_{\geq m}(s)$ of the connected uniform random tree of $10^4$ vertices. For clarity, only the highest 30 values of $m$ are plotted. This tree was generated by the Aldous--Broder algorithm \cite{aldous1990random,broder1989generating} which we run  on the complete graph of $10^4$ vertices. 
}
\label{f3}
\end{figure*}

Figure~\ref{f3} shows the cumulative distribution $P^\text{(cum)}_{\geq m}(s)$ for a quite different tree, namely, for a connected uniform random tree, whose Hausdorff dimension equals $2$, that is, this random tree is a ``large world''. 
The figure demonstrates that the cumulative distribution does not have a power-law asymptotics.


\begin{figure*}[t]
\begin{flushleft}
\includegraphics[angle=0,width=225pt]{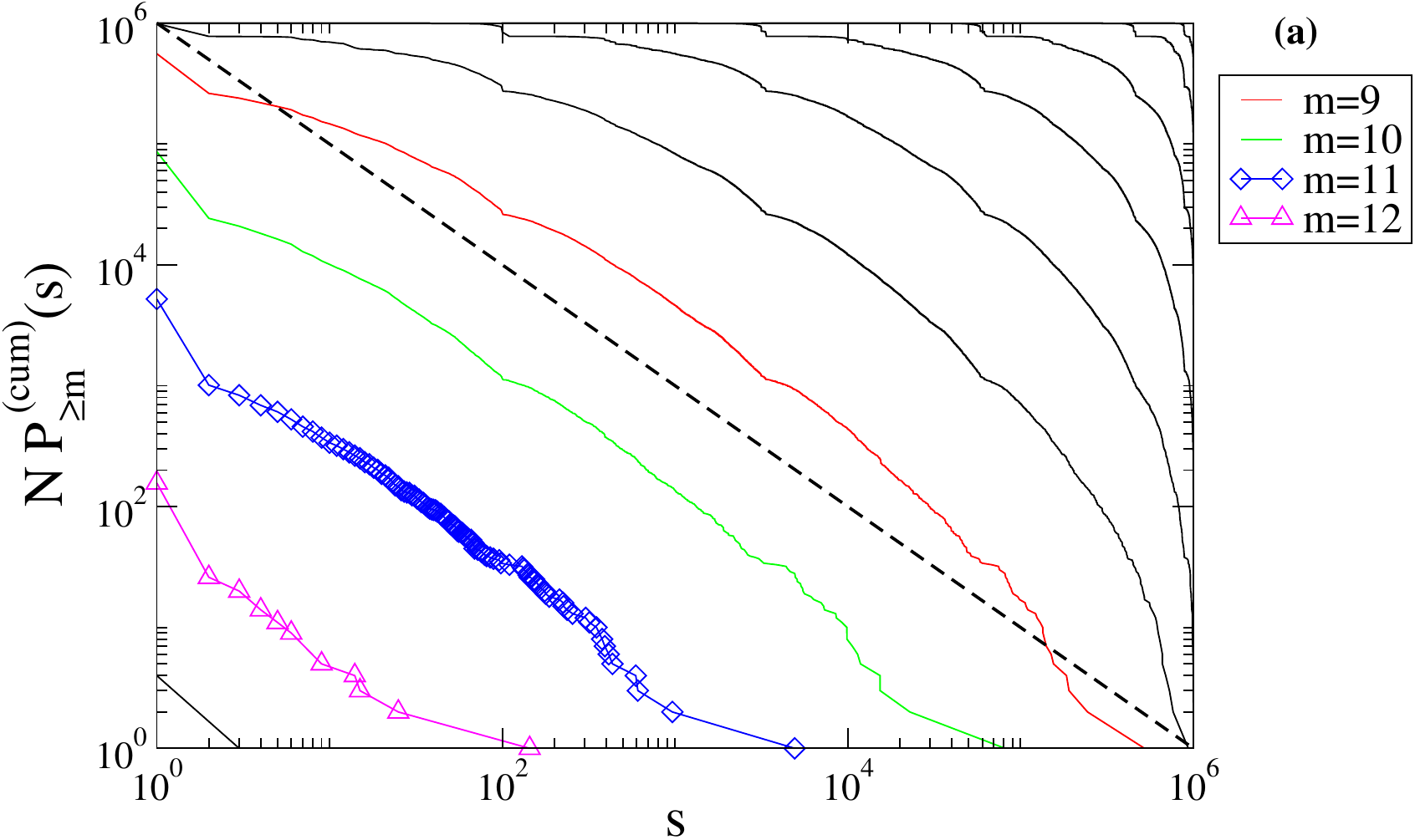} 
\ 
\includegraphics[angle=0,width=225pt]{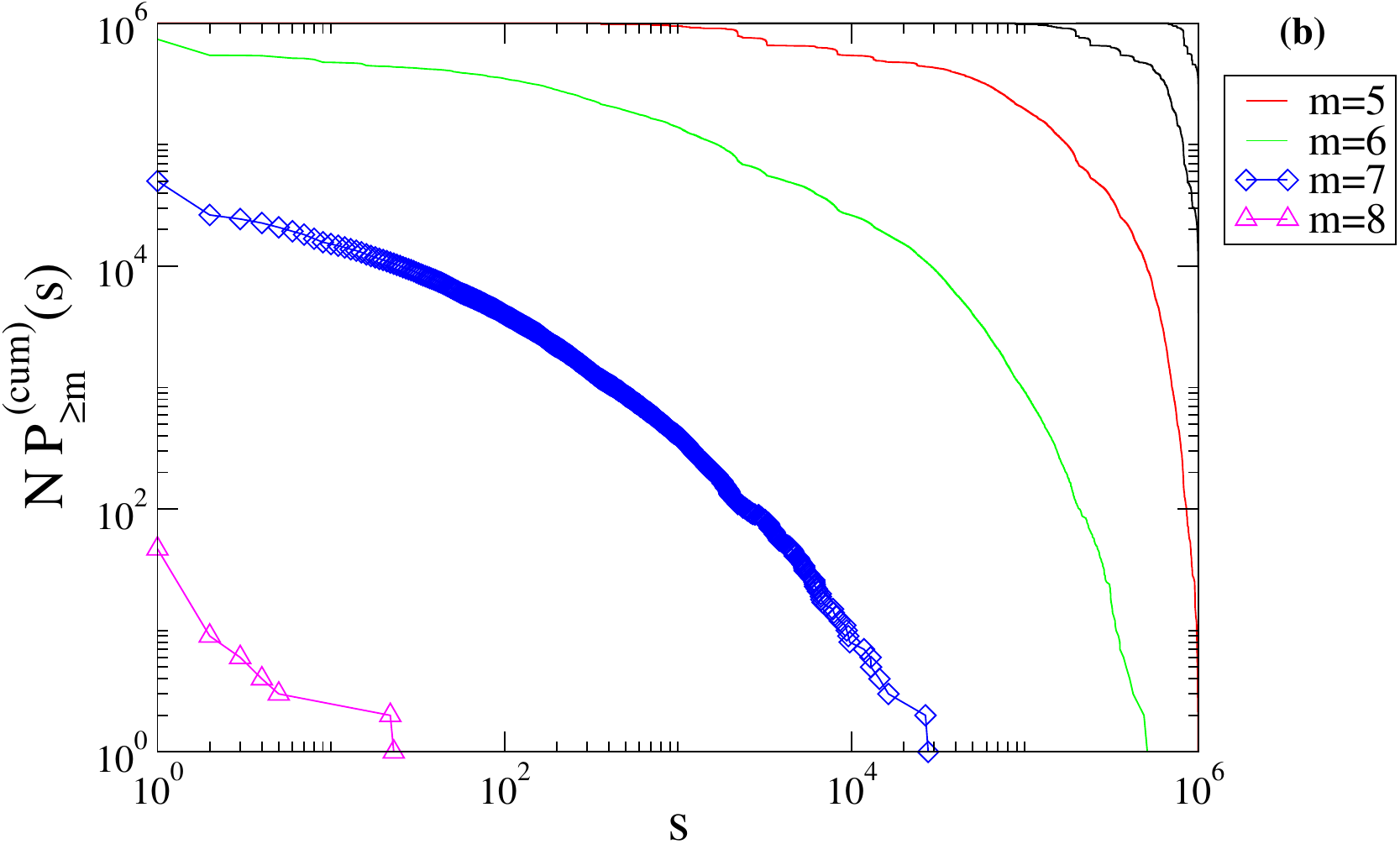}
\\[3pt]
\includegraphics[angle=0,width=225pt]{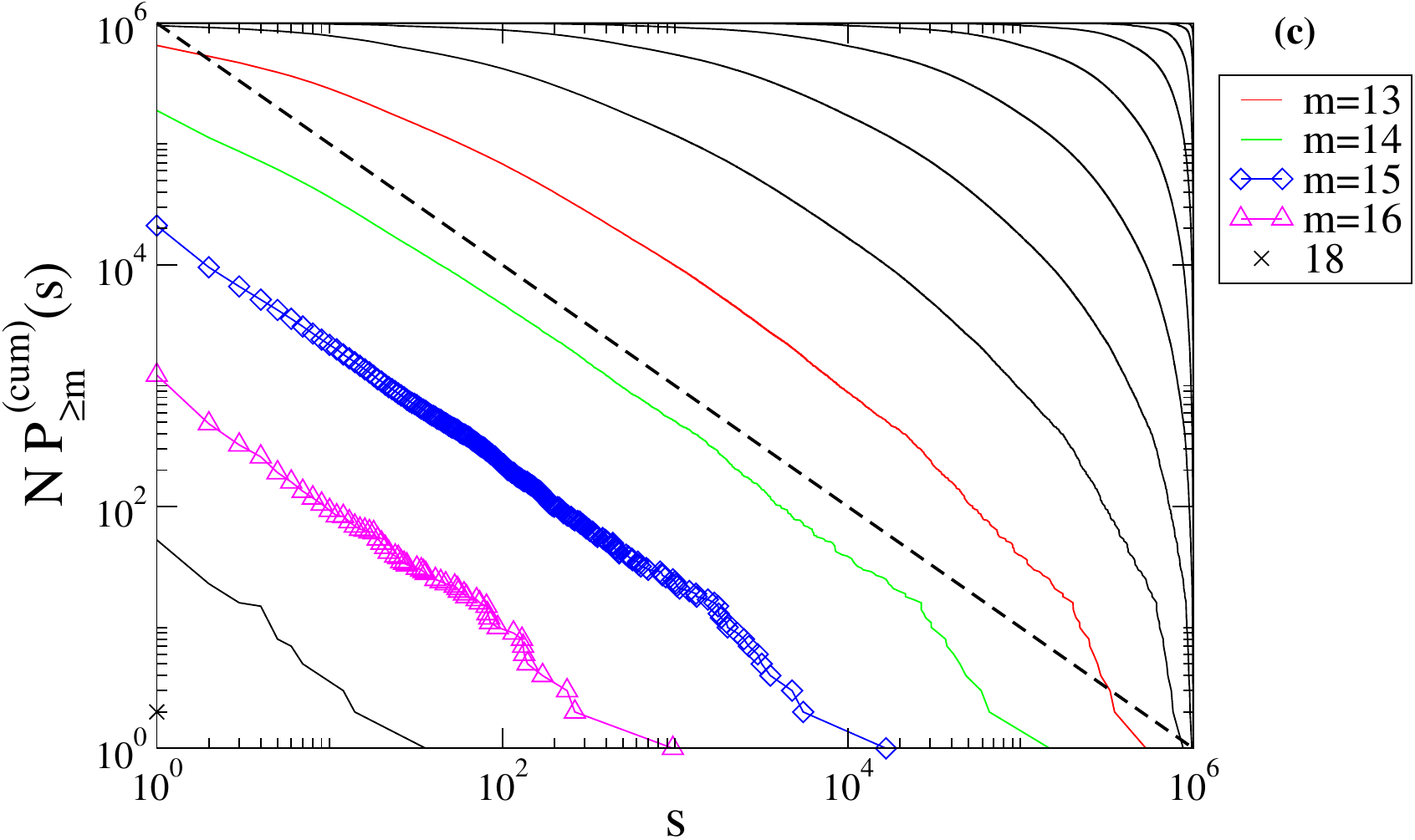}
\ 
\includegraphics[angle=0,width=225pt]{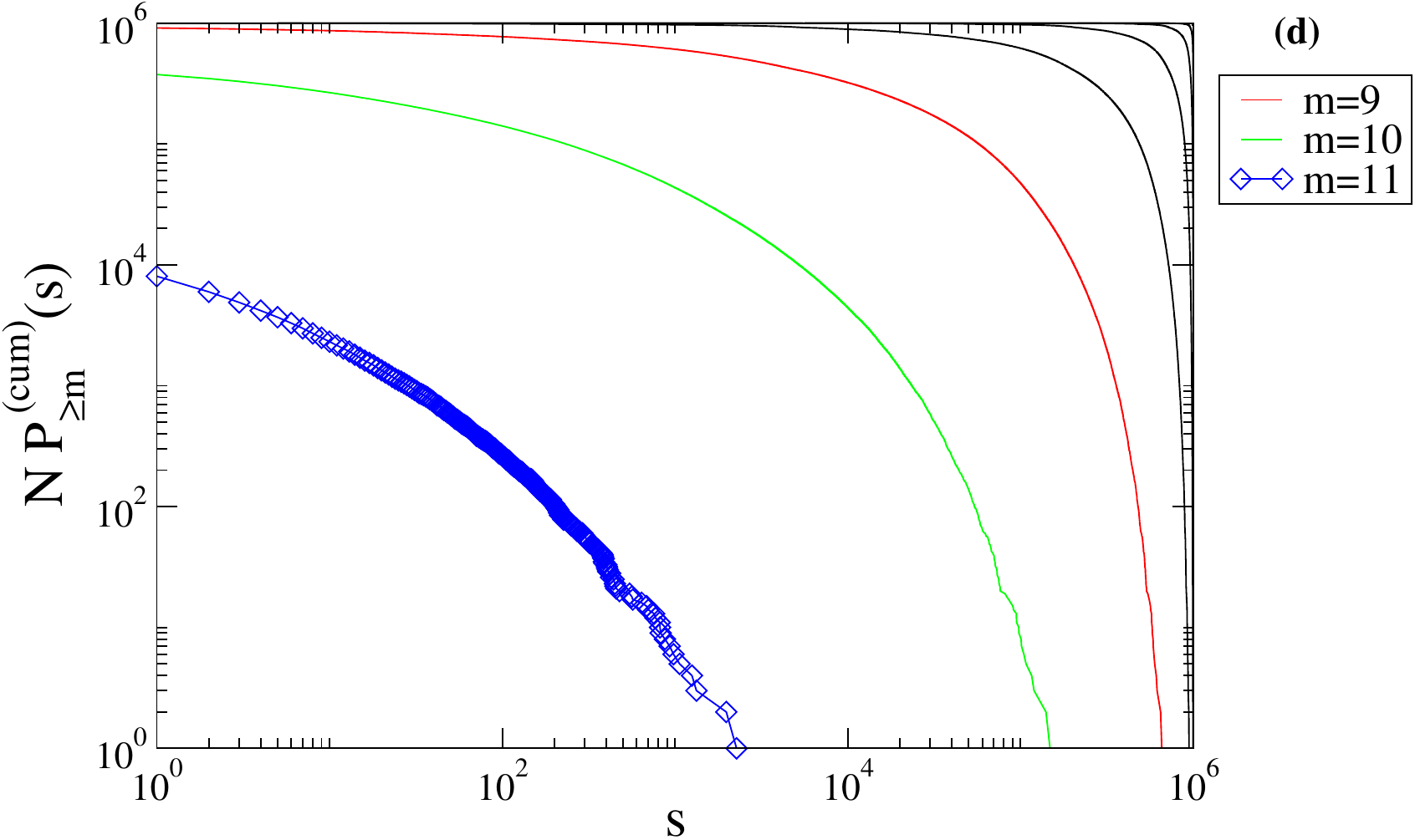}
\\[3pt]
\includegraphics[angle=0,width=225pt]{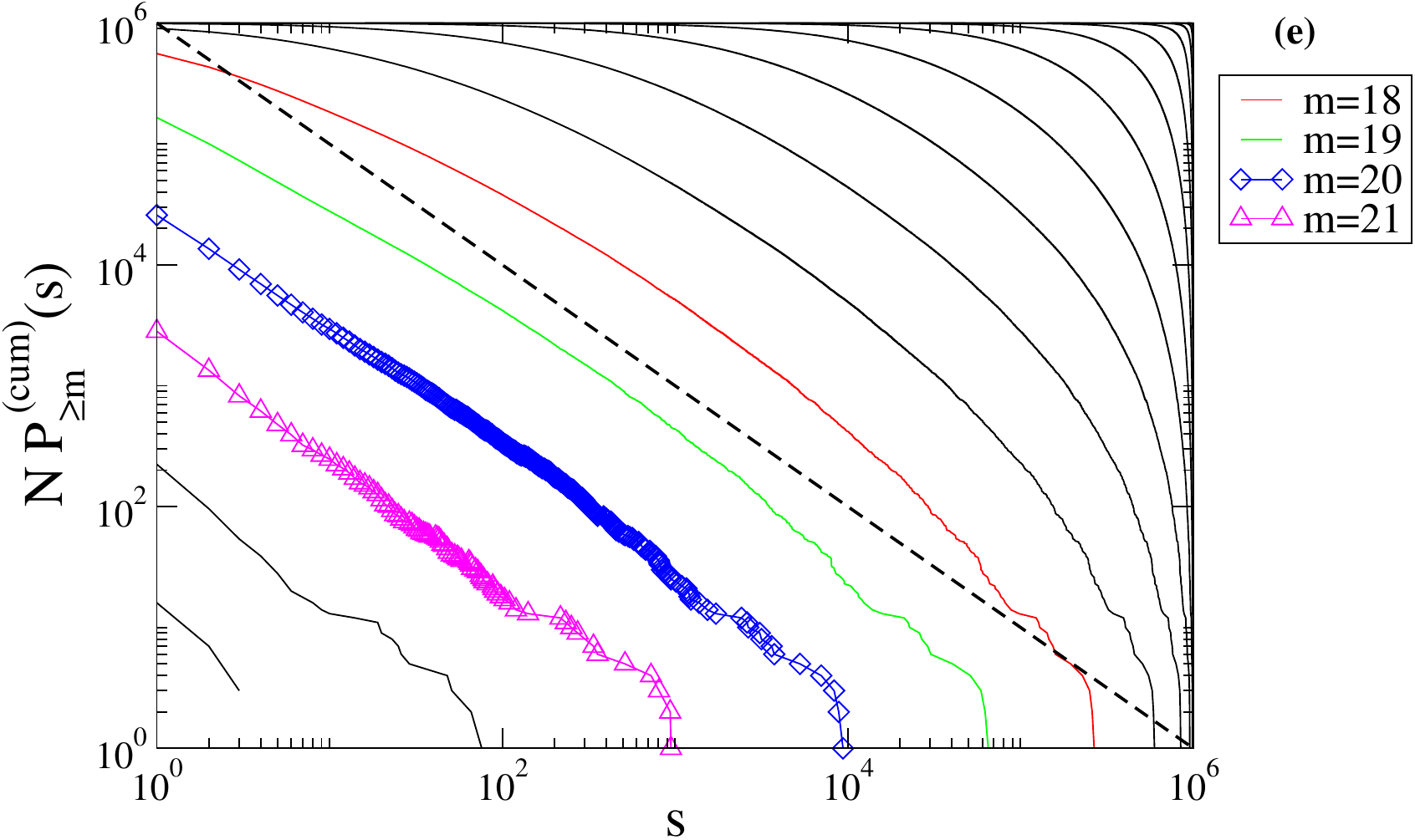} 
\ 
\includegraphics[angle=0,width=225pt]{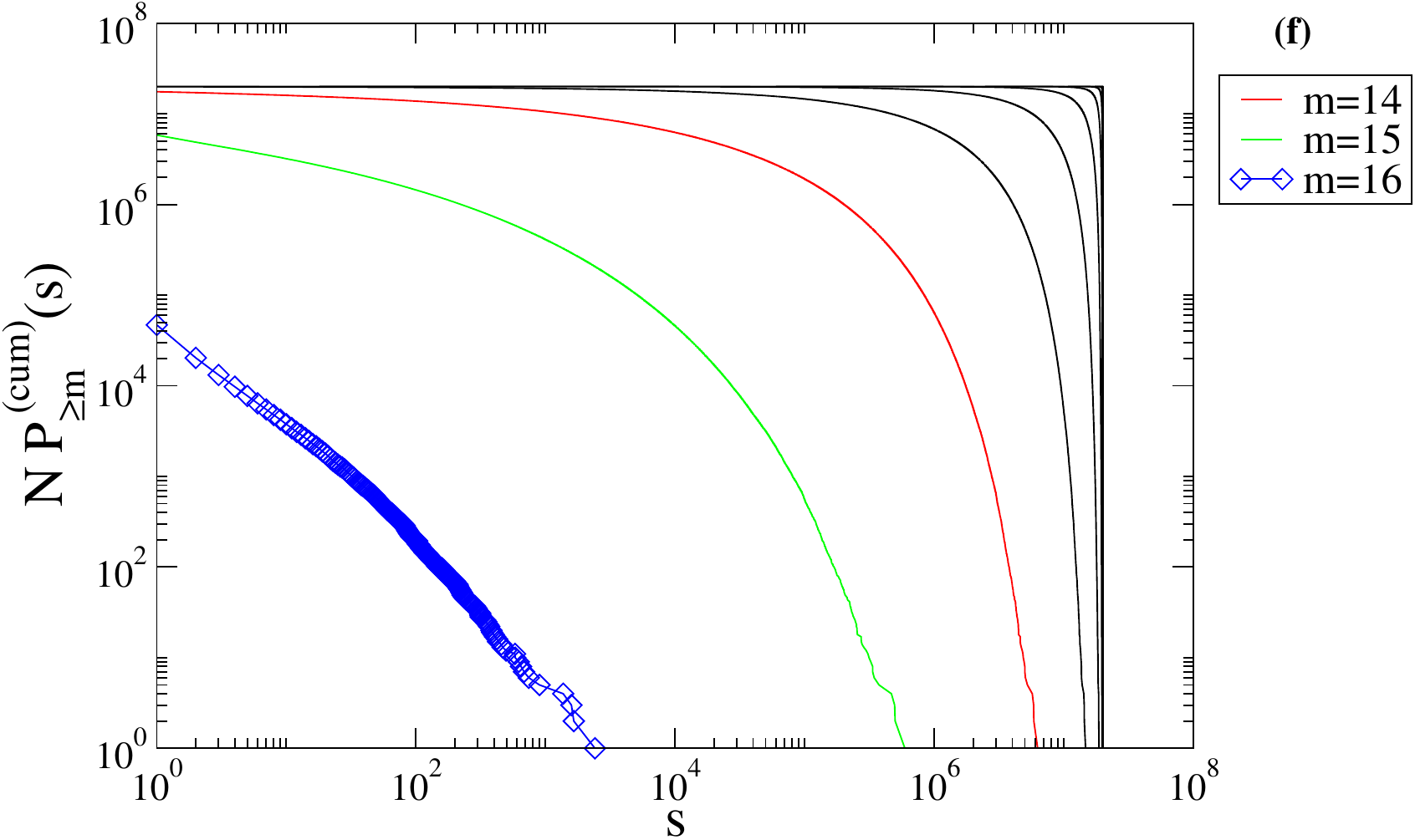}
\end{flushleft}
\caption{Cumulative distribution $P^\text{(cum)}_{\geq m}(s)$ of six random growing networks. Dashed lines have slope $-1$.
(a,c,e) Each new vertex in a recursive network attaches, with equal probability, to one or two existing vertices selected by the same attachment rules as for the recursive trees in Fig.~\ref{f2}(a,b,c), respectively. (b,d,f) Each new vertex in a recursive network attaches to two existing vertices selected by the same attachment rules as for the recursive trees in Fig.~\ref{f2}(a,b,c), respectively. The networks in (a)--(e) contain $10^6$ vertices, the network in (f) has $2\times10^7$ vertices.
}
\label{f4}
\end{figure*}


Figure~\ref{f4} shows the cumulative distributions $P^\text{(cum)}_{\geq m}(s)$ for loopy recursive networks whose growth is similar to the recursive trees in Fig.~\ref{f2} with one difference. In contrast to the recursive trees, each new vertex in the networks in Fig.~\ref{f4}(a,c,e) attaches, with equal probability, to one or two existing vertices, and each new vertex in the networks in Fig.~\ref{f4}(b,d,f) attaches to two existing vertices. The existing vertices for attachment are chosen by the rules implemented for the recursive trees in Fig.~1(a,b,c). The degree distributions and correlations of the trees in Fig.~1(a,b,c) and the loopy networks in, respectively, Fig.~\ref{f4}(a,c,e) and Fig.~\ref{f4}(b,d,f) are similar. 
One can see the asymptotics $1/s$ of the cumulative distributions $P^\text{(cum)}_{\geq m}(s)$ for the loopy growing networks in Fig.~\ref{f4}(a,c,e), which have vertices of degree $1$, while this asymptotics is not observed in the loopy growing networks in Fig.~\ref{f4}(b,d,f), which have no vertices of degree $1$, despite their large sizes. 

In Table~\ref{t1} we list the basic structural characteristics of the synthetic networks considered in this paper. 

\begin{table*}[h]
	\centering
	\begin{tabular}{|c|c|c|c|c|c|}
	  \hline
	  Network & (Fig.) & $N$ & $k_{max}$ & $\ell$ & $\ell_{max}$ \\
	  \hline
	  \hline
	  \hline
	  \hline
	  Erd\H{o}s-R\'enyi & (\ref{f1}) & $10^6$ & 20 & 8.76 & 17 \\
	  \hline
	  \hline
	  \hline
	  Recursive random 2.2 tree & (\ref{f2}a) & $10^6$ & 78,743 & 6.24 & 24 \\
	  \hline
	  Recursive BA tree & (\ref{f2}b) & $10^5$ & 514 & 12.45 & 32 \\
	  \hline
	  Recursive random tree & (\ref{f2}c) & $10^5$ & 16 & 20.54 & 51 \\
	  \hline
	  \hline
	  \hline
	  Uniform random tree & (\ref{f3}) & $10^4$ & 8 & 106.2 & 285 \\
	  \hline
	  \hline
	  \hline
	  Recursive 2.2 mixed & (\ref{f4}a) & $10^6$ & 119,422 & 4.173 & 13 \\
	  \hline
	  Recursive 2.2 to two & (\ref{f4}b) & $10^6$ & 178,791 & 3.435 & 8 \\
	  \hline
	  \hline
	  Recursive BA mixed & (\ref{f4}c) & $10^6$ & 2,241 & 7.66 & 18 \\
	  \hline
	  Recursive BA to two & (\ref{f4}d) & $10^6$ & 2,647 & 6.73 & 12 \\
	  \hline
	  \hline
	  Recursive random mixed & (\ref{f4}e) & $10^6$ & 32 & 11.24 & 23 \\
	  \hline
	  Recursive random to two & (\ref{f4}f) & $2\times10^7$ & 43 & 10.89 & 16 \\
	  \hline
	 \end{tabular}
\caption{Basic structural characteristics of the synthetic networks considered: each line has information specifying a network realization, the figure where numerical results for the distributions are plotted, the size $N$ of the largest component, the maximum degree $k_{max}$, the average path length $\ell$, and the maximum path length $\ell_{max}$. 
}	 
 \label{t1}
\end{table*}




\section{The statistics of remote regions in real-world networks}
\label{s3}

\begin{figure*}[t]
\begin{flushleft}
\includegraphics[angle=0,width=225pt]{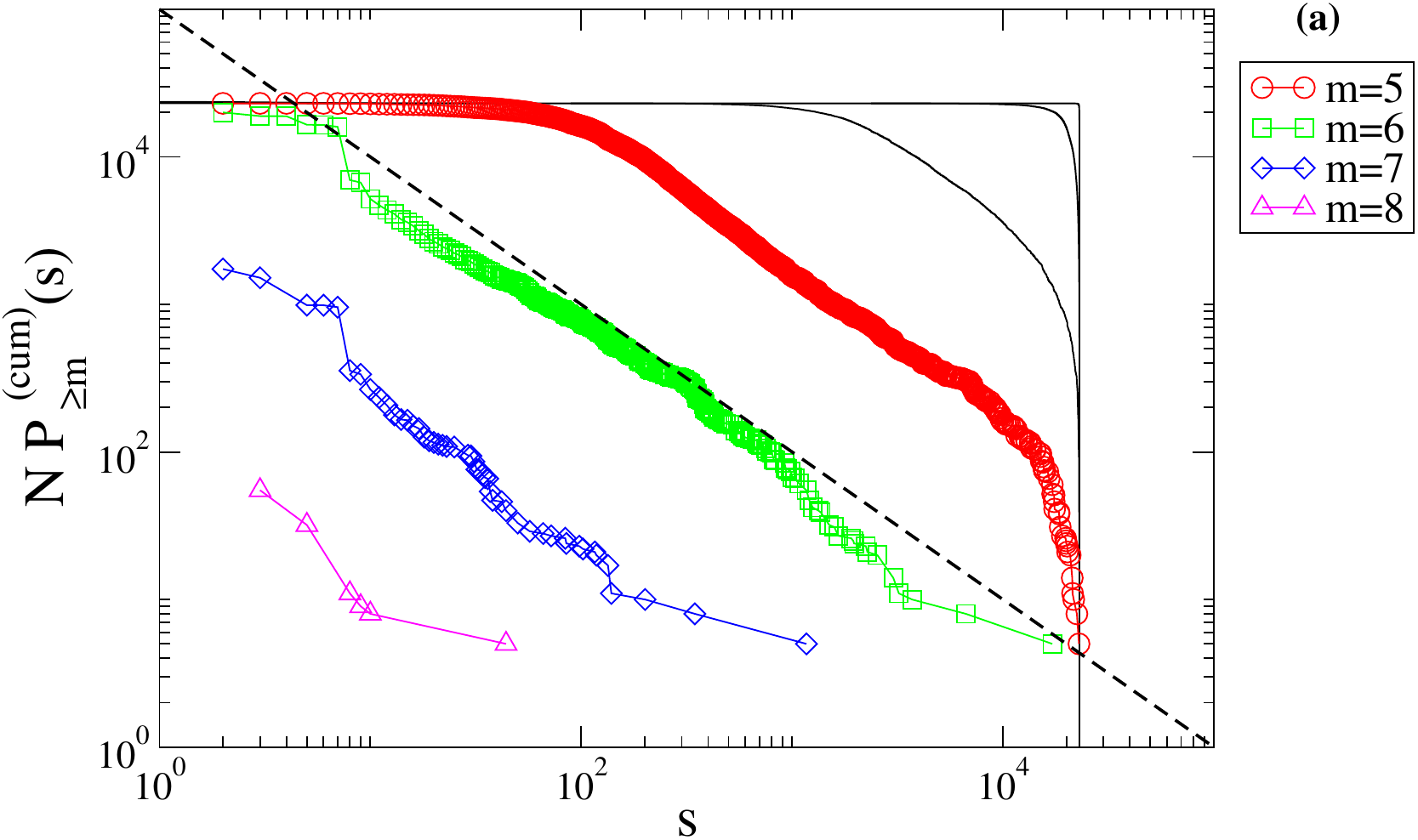}
\ 
\includegraphics[angle=0,width=225pt]{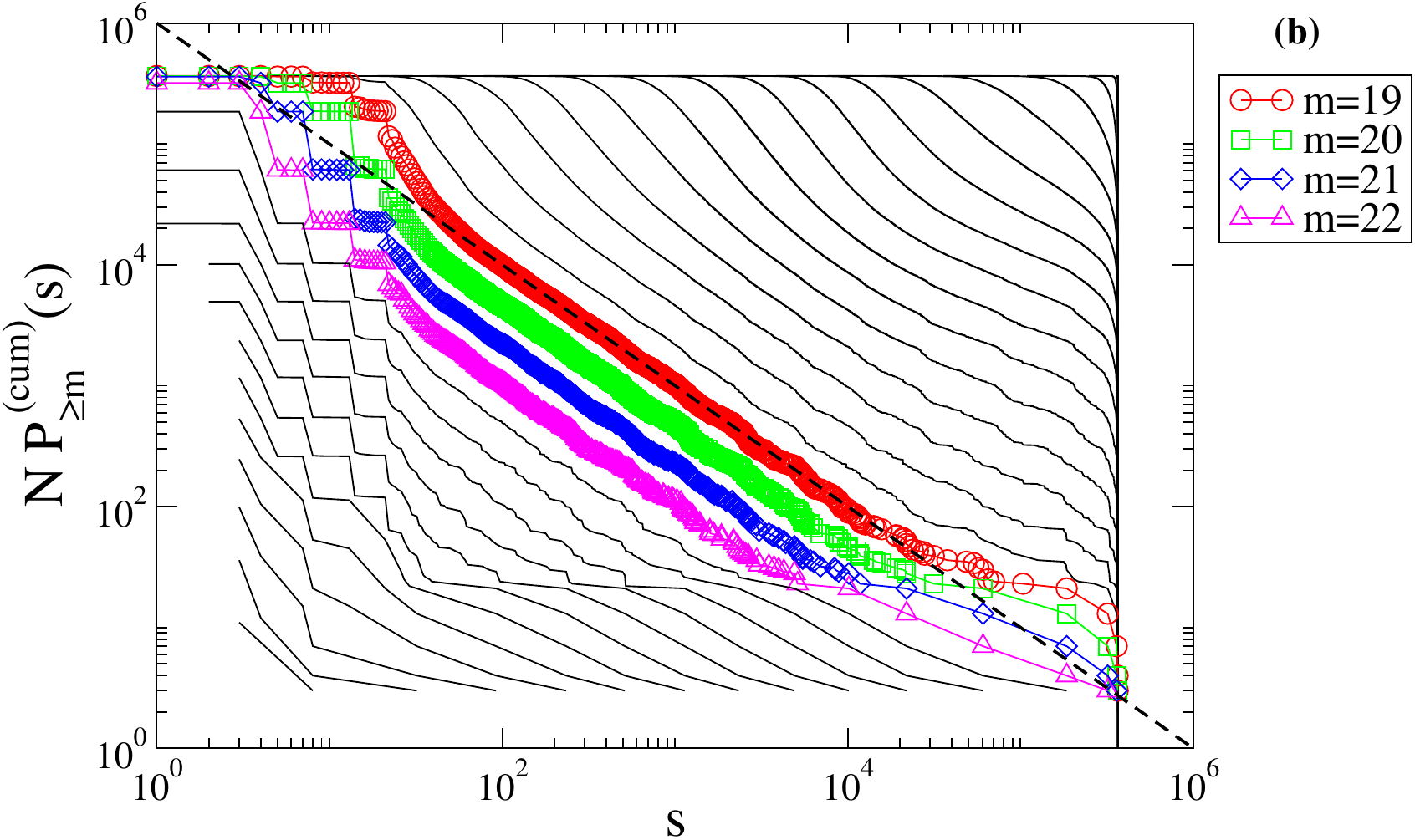}
\\[3pt]
\includegraphics[angle=0,width=225pt]{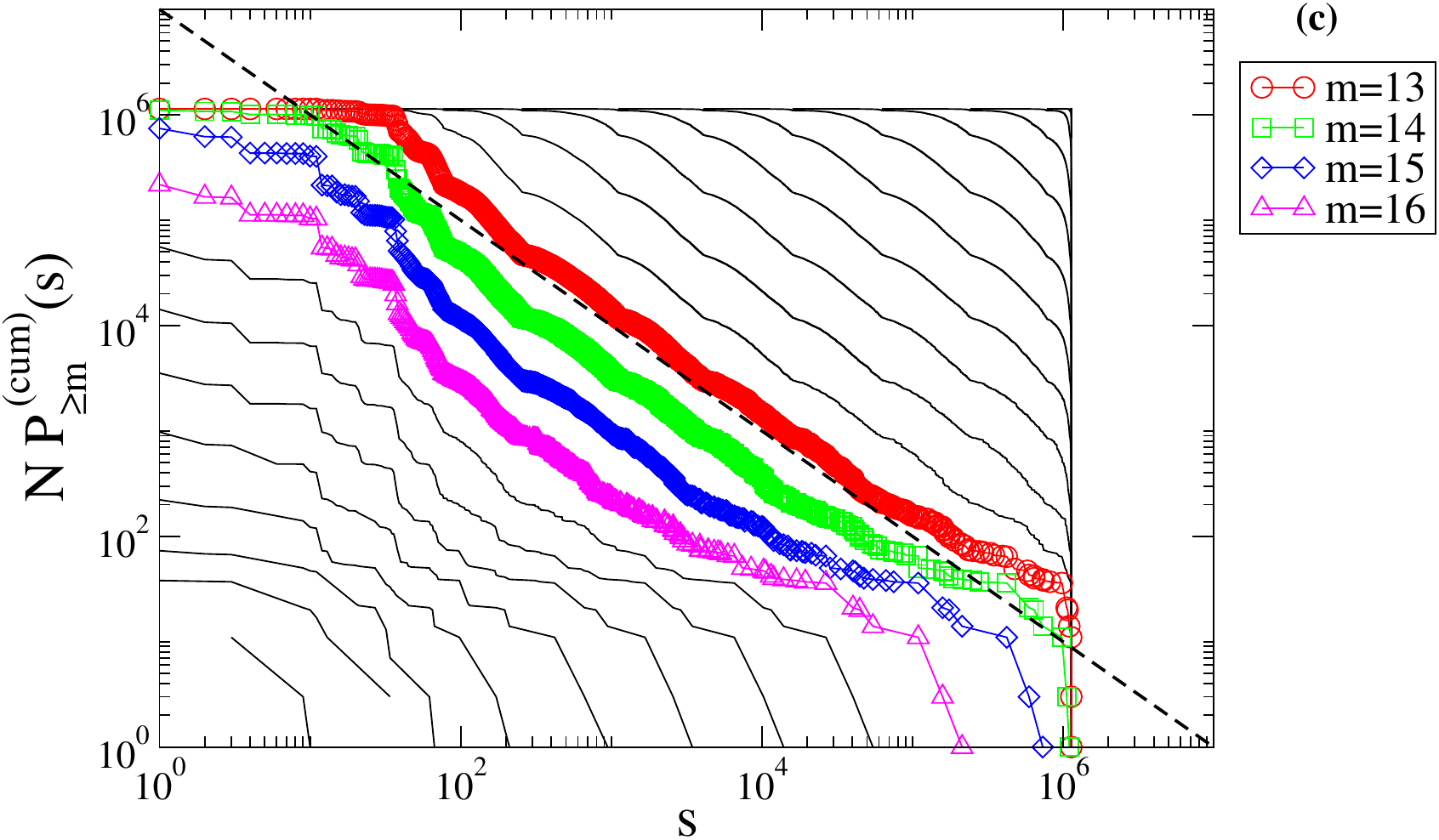}
\ 
\includegraphics[angle=0,width=225pt]{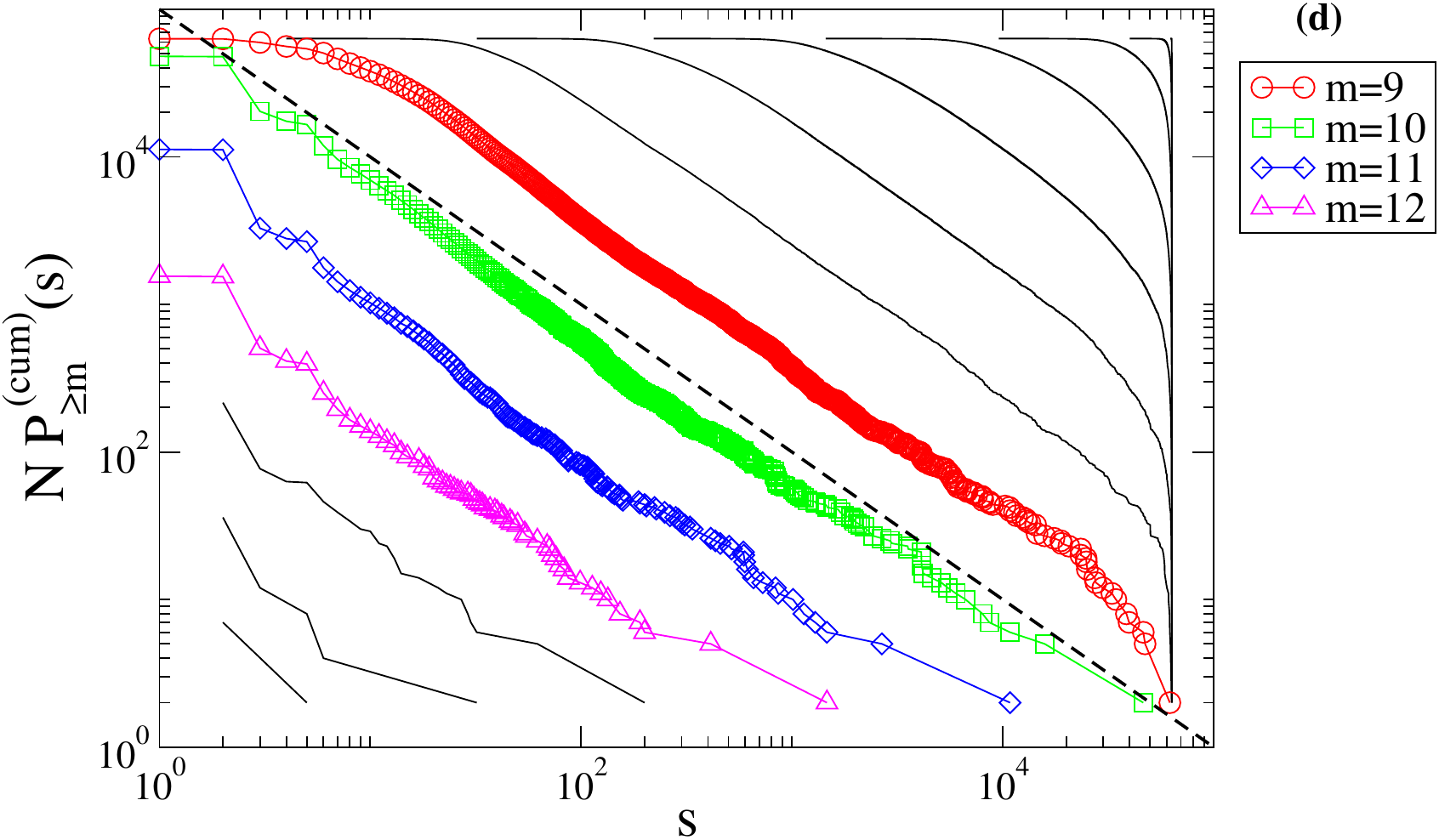}
\end{flushleft}
\caption{The cumulative distributions $P^\text{(cum)}_{\geq m}(s)$ for the four large collaboration social networks. Dashed lines have slope $-1$. (a) The FP5 net of $25,287$ vertices \cite{almendral2007675,almendral2007TheIO}. (b) The largest component of CiteSeer with $365,154$ vertices \cite{Kunegis2013KONECTTK}. (c) A documented region of the Youtube friends network of $1,134,890$ vertices \cite{mislove-2007-socialnetworks}. (d) A large component in Facebook with $63,394$ vertices \cite{viswanath-2009-activity}. 
}
\label{f5}
\end{figure*}

Real-world networks typically have more complicated architectures than synthetic ones, and so one could expect that the observation of the inverse square law in real networks is more difficult. Surprisingly, this is not the case.  
In Table~\ref{t2} we list the basic structural characteristics of the real-world networks considered in this paper. 
Figure~\ref{f5} shows the cumulative distributions $P^\text{(cum)}_{\geq m}(s)$ for the maps of the large regions of four collaboration and social networks, namely, the FP5 net, CiteSeer, the Youtube friends network, and Facebook. 
For all four sets of cumulative distributions we observe the $1/s$ asymptotics. 


\begin{table*}[t]
	\centering
	\begin{tabular}{|c|c|c|c|c|c|}
	  \hline
	  Network & (Fig.) & $N$ & $k_{max}$ & $\ell$ & $\ell_{max}$ \\
	  \hline
	  \hline
	  \hline
	  \hline
	  FP5 & (\ref{f5}a) & 25,287 & 2,783 & 3.14 & 8 \\
	  \hline
	  CiteSeer & (\ref{f5}b) & 365,154 & 1,739 & 6.470 & 34 \\
	  \hline
	  YouTube & (\ref{f5}c) & 1,134,890 & 28,754 & 5.279 & 24 \\
	  \hline
	  Facebook & (\ref{f5}d) & 63,392 & 1,098 & 4.322 & 15 \\
	  \hline
	  \hline
	  \hline
	  Routers CAIDA & (\ref{f6}a) & 192,244 & 1,071 & 6.98 & 26 \\
	 \hline
	  AS CAIDA & (\ref{f6}b) & 26,475 & 2628 & 3.876 & 17 \\
	  \hline
	  \hline
	  \hline
	  US power grid & (\ref{f7}a) & 4,941 & 19 & 18.99 & 46 \\
	  \hline
	  Road newtork PA & (\ref{f7}b) & 1,087,562 & 9 & 308.0 & 794 \\
	  \hline
	  \hline
	  \hline
	  Google web & (\ref{f8}a) & 15,763 & 11,401 & 2.517 & 7 \\
	  \hline
	  \hline
	  \hline
	  Web Stanford & (\ref{f9}) & 255,265 & 38,625 & 6.815 & 164 \\
	  \hline
	 \end{tabular}
\caption{Basic structural characteristics of the real-world networks considered: each line has information specifying a network, the figure where numerical results for the distributions are plotted, the size $N$ of the largest component, the maximum degree $k_{max}$, the average path length $\ell$, and the maximum path length $\ell_{max}$.
}	 
 \label{t2}
\end{table*}



\begin{figure*}[t]
\begin{flushleft}
\includegraphics[angle=0,width=225pt]{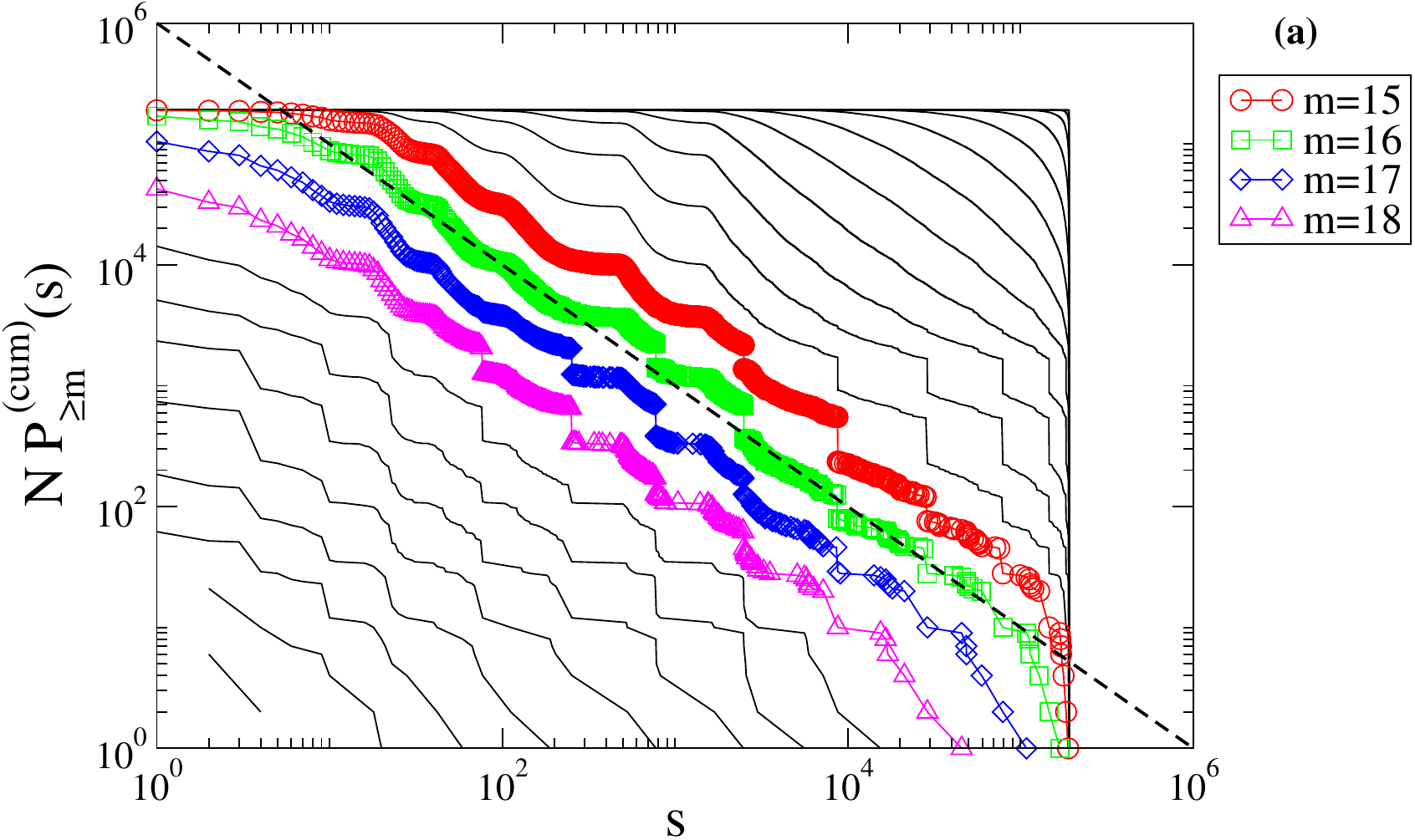}
\ 
\includegraphics[angle=0,width=225pt]{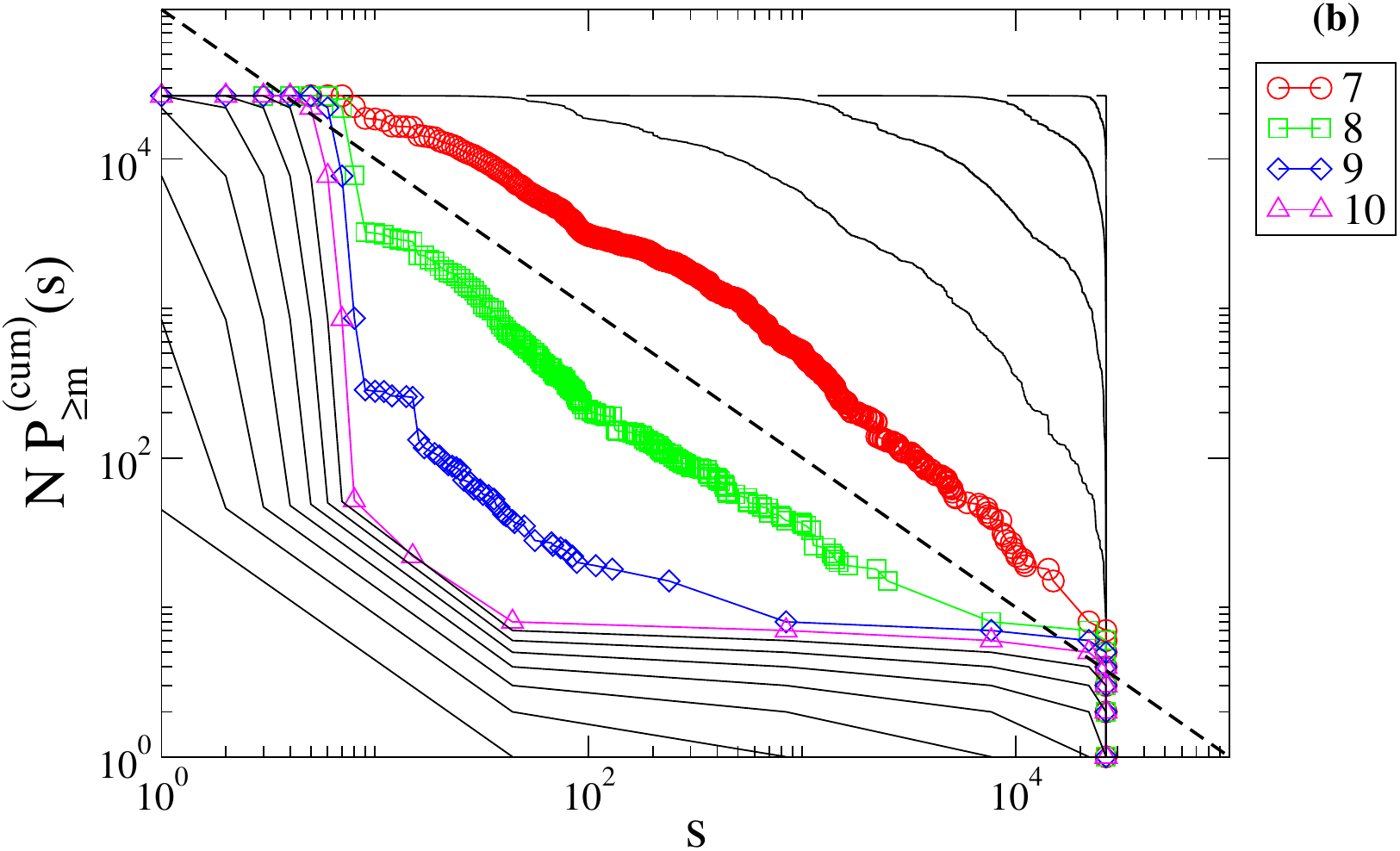}
\end{flushleft}
\caption{The cumulative distributions $P^\text{(cum)}_{\geq m}(s)$ for the Internet networks. Dashed lines have slope $-1$. (a) The CAIDA map of the routers in the Internet including $192,244$ routers \cite{caida_router}. (b) The CAIDA map of the Autonomous Systems (AS) with $26,475$ AS. \cite{as_caida}. 
}
\label{f6}
\end{figure*}

\begin{figure*}[t]
\begin{flushleft}
\includegraphics[angle=0,width=225pt]{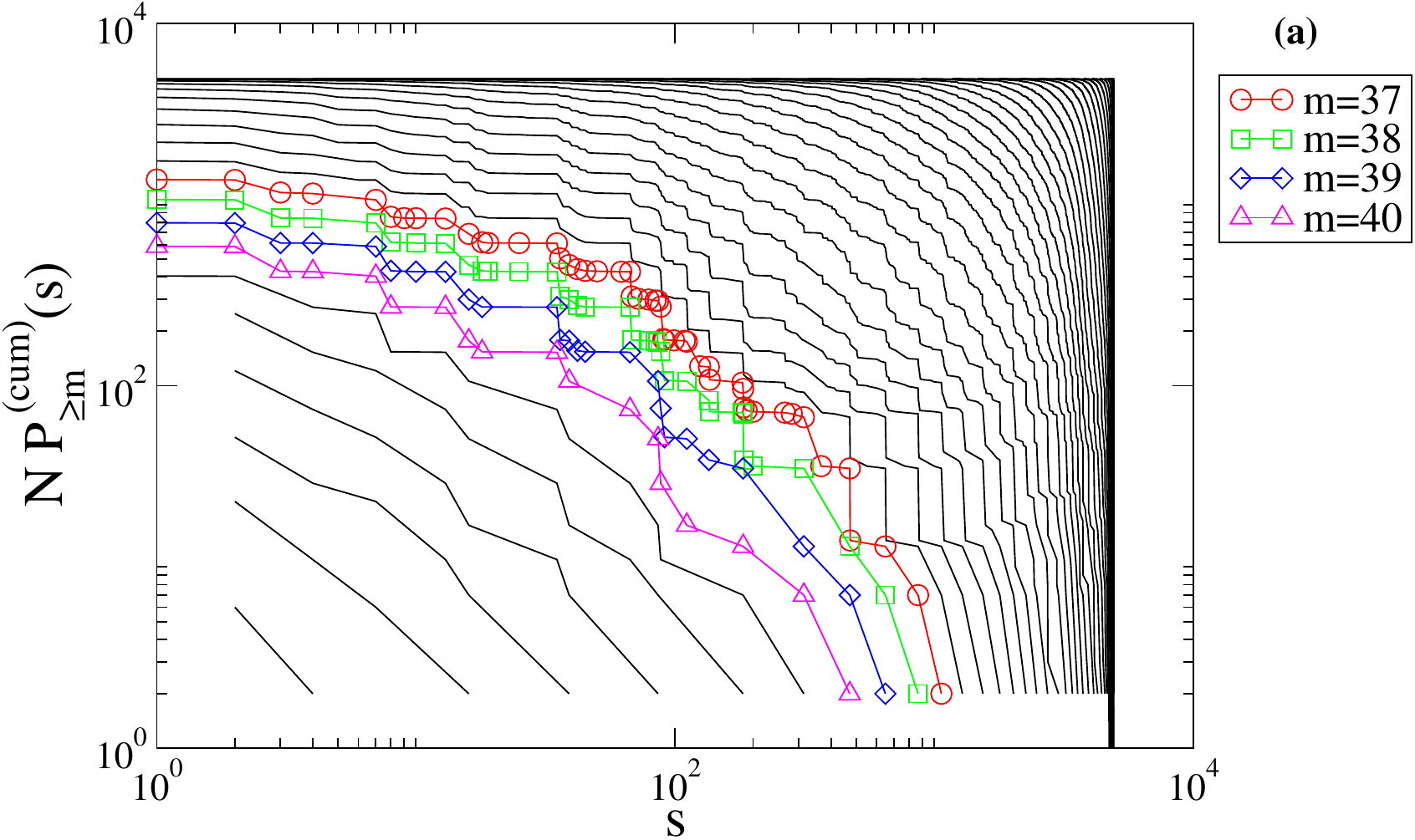}
\ 
\includegraphics[angle=0,width=225pt]{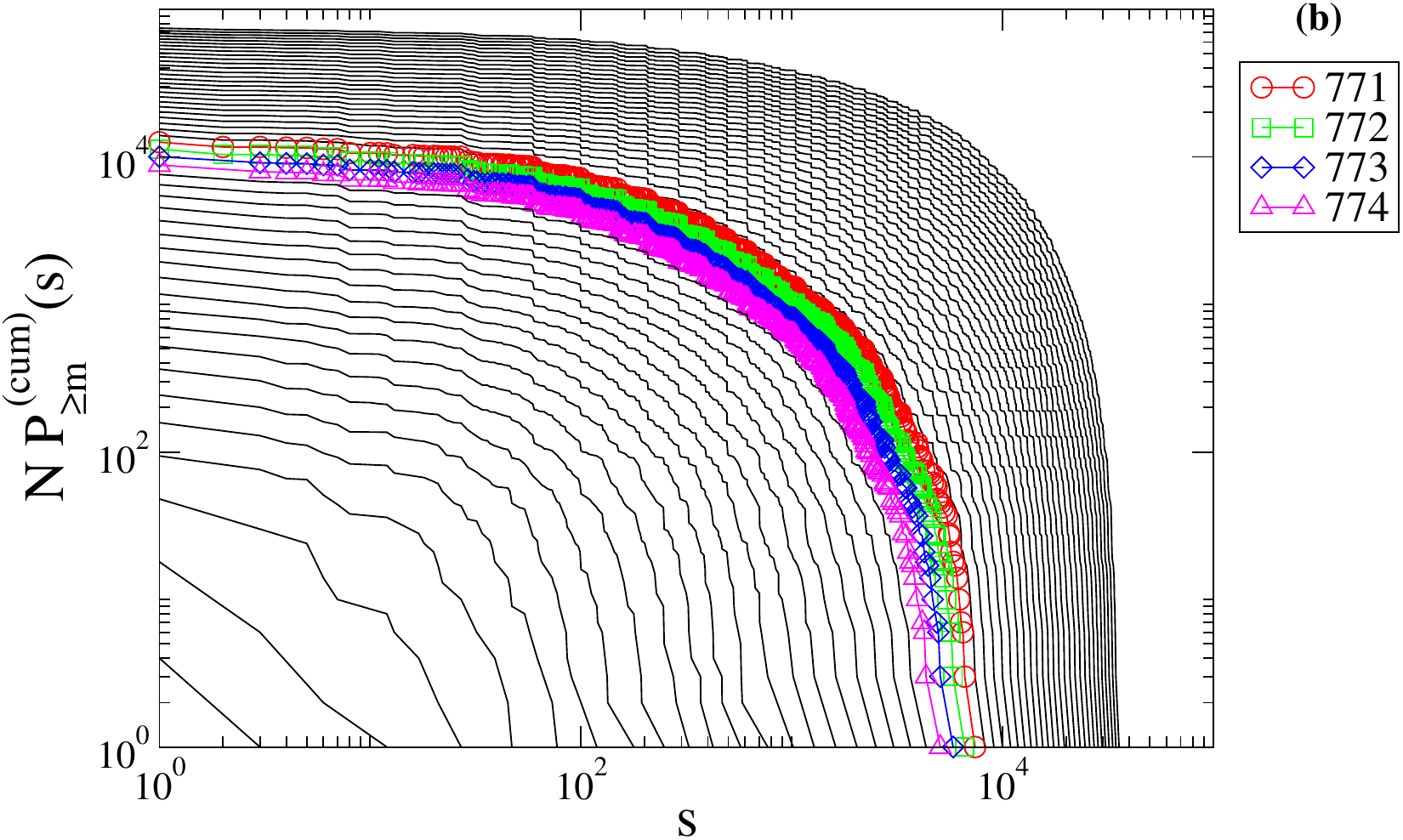}
\end{flushleft}
\caption{The cumulative distributions $P^\text{(cum)}_{\geq m}(s)$ for (a) the US power grid of $4,941$ vertices \cite{Watts1998-ve} and (b) the US road network Pennsylvania with $1,087,562$ vertices \cite{Leskovec2009-vs}. For clarity, in (b), only the highest 50 values of $m$ are plotted.
}
\label{f7}
\end{figure*}

We also observe this asymptotics inspecting the cumulative distributions $P^\text{(cum)}_{\geq m}(s)$ for the Internet networks \cite{pastor2004evolution}: the maps of the routers and the autonomous systems, see Fig.~\ref{f6}. On the other hand, as is natural, the US power grid and the road network of Pennsylvania, which are two-dimensional networks, do not demonstrate the power-law asymptotics of $P^\text{(cum)}_{\geq m}(s)$, see Fig.~\ref{f7}. 


\begin{figure*}[t]
\begin{flushleft}
\includegraphics[angle=0,width=225pt]{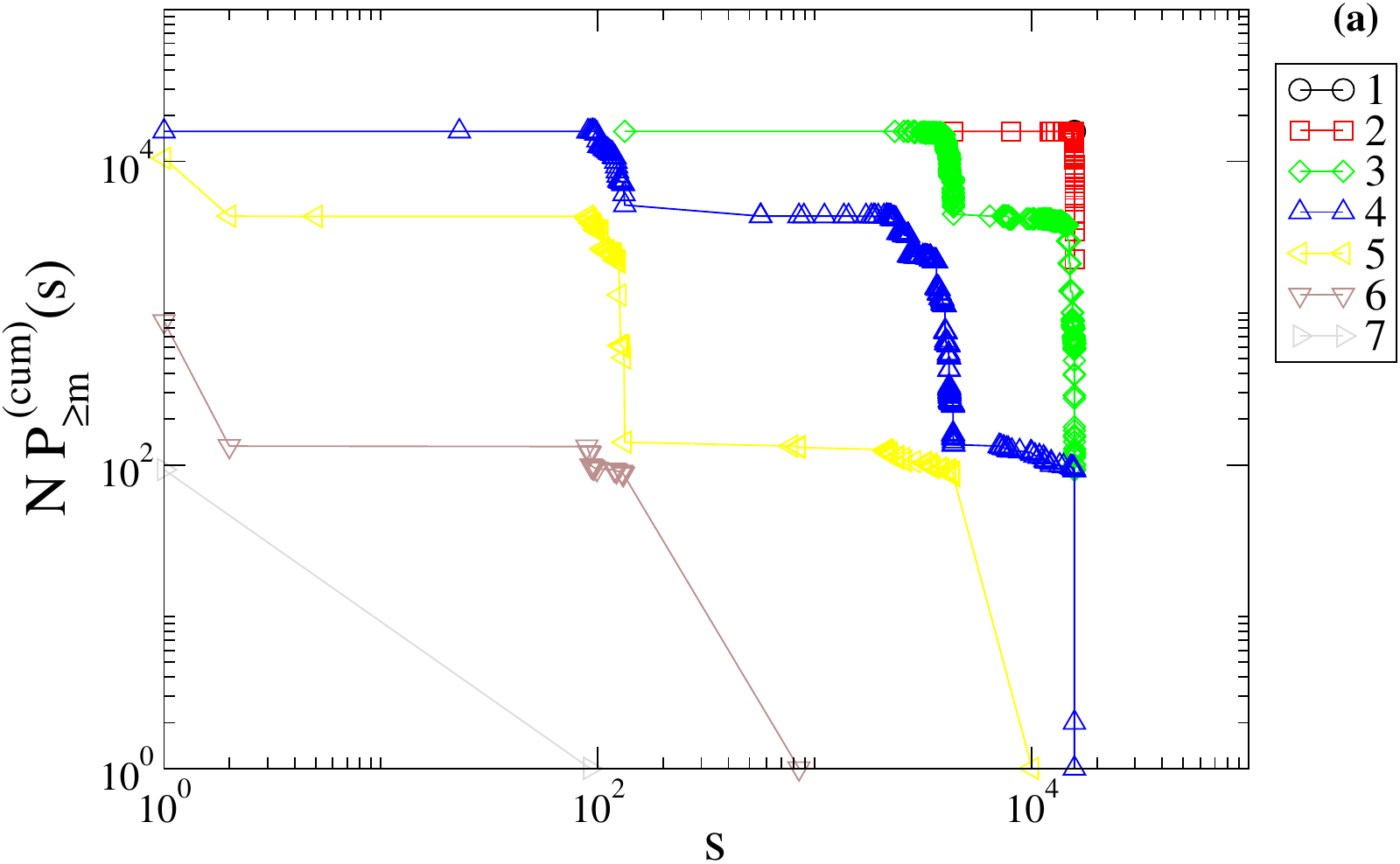}
\ 
\includegraphics[angle=0,width=238pt]{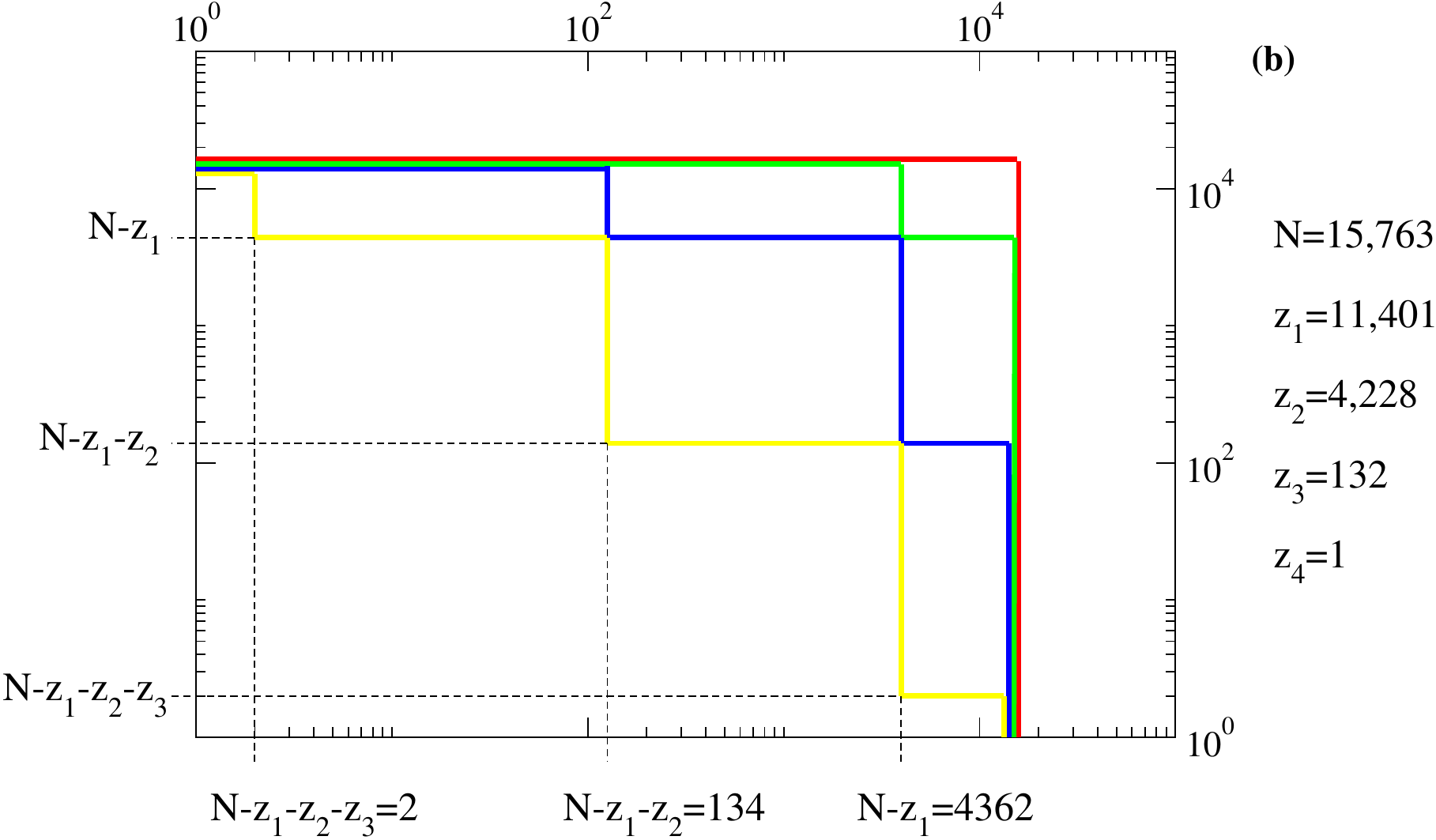}
\end{flushleft}
\caption{(a) The cumulative distributions $P^\text{(cum)}_{\geq m}(s)$ of the undirected projection of a network (15,763 vertices and 171,206 edges) of hyperlinks between pages within Google's sites \cite{palla2007directed}. (b) The theoretical cumulative distributions $P^\text{(cum)}_{\geq m}(s)$ of the model tree-like network mimicking the Google net: it has the hub with the same numbers of the first-, second-, third-, and fourth-nearest neighbours, $z_1=11,401$, $z_2=4228$, $z_3=132$, and $z_4=1$, as the Google net.
}
\label{f8}
\end{figure*}


Figure~\ref{f8}(a) shows the cumulative distributions $P^\text{(cum)}_{\geq m}(s)$ for a real-world network with a very large hub. This is the undirected projection of a network of 171,206 hyperlinks between 15,763 pages within Google's sites. The largest hub in this network has huge degree 11,401, which shapes the architecture of this specific network. 
The steps in the empirical cumulative distributions in Fig.~\ref{f8}(a) can be reproduced in a tree-like model network mimicking the structure of the Google net. Imagine a tree-like network with the hub having the same numbers of the first-, second-, third-, etc.-nearest neighbours as the hub in the Google net. For this model network one can easily estimate $P^\text{(cum)}_{\geq m}(s)$, see Fig.~\ref{f8}(b), and get a quantitative agreement with the empirical distribution for $m=2$, $3$, $4$, and $5$. The small size of this network does not allow us to check whether this special network architecture still provides the inverse square law or not.  


\begin{figure*}[t]
\begin{flushleft}
\includegraphics[angle=0,width=225pt]{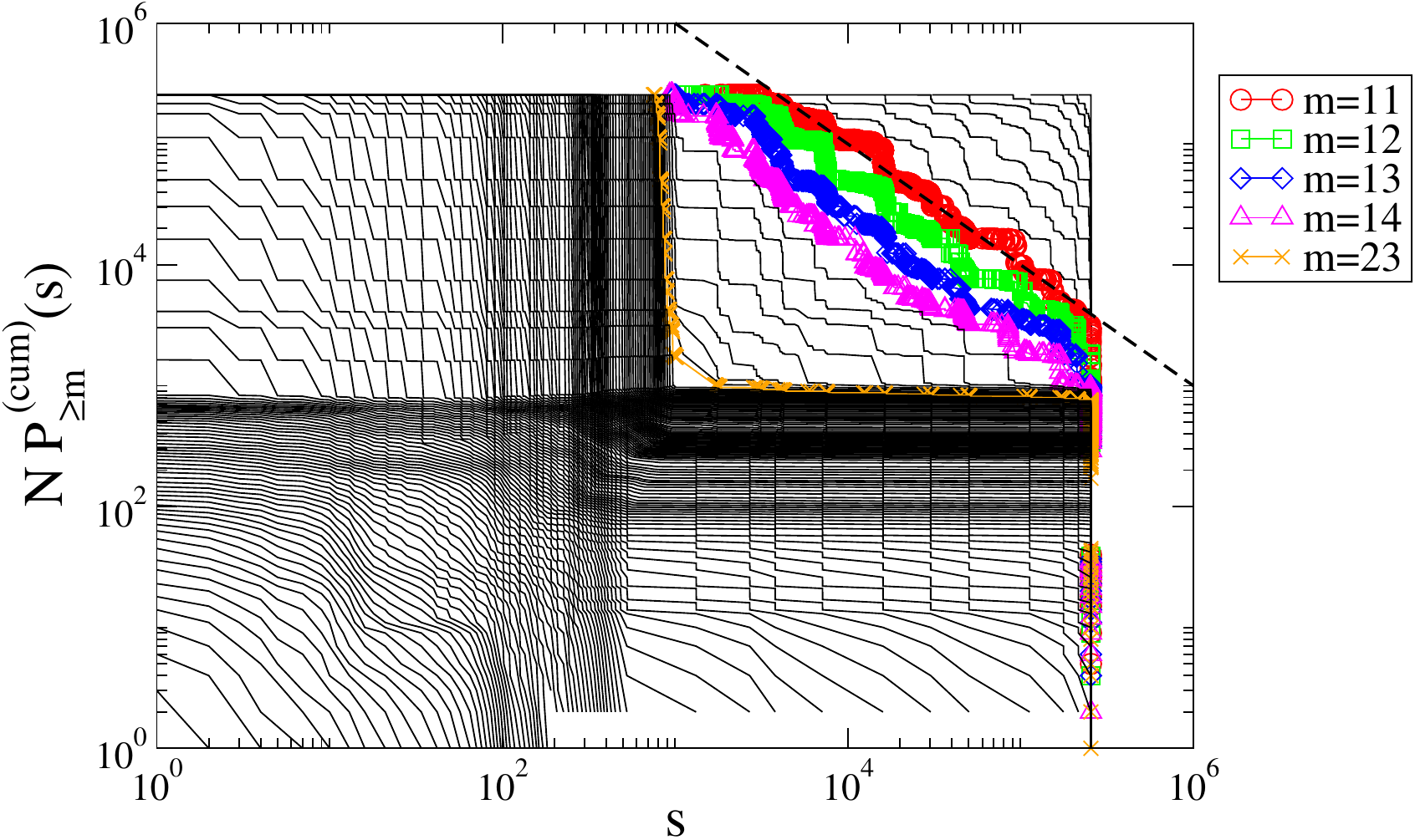}
\end{flushleft}
\caption{The cumulative distributions $P^\text{(cum)}_{\geq m}(s)$ for the undirected projection of a large Stanford Web domain containing $255,265$ vertices \cite{Leskovec2009-vs}. Dashed line has slope $-1$.
}
\label{f9}
\end{figure*}

\begin{figure*}[t]
\begin{flushleft}
\includegraphics[angle=0,width=2\columnwidth]{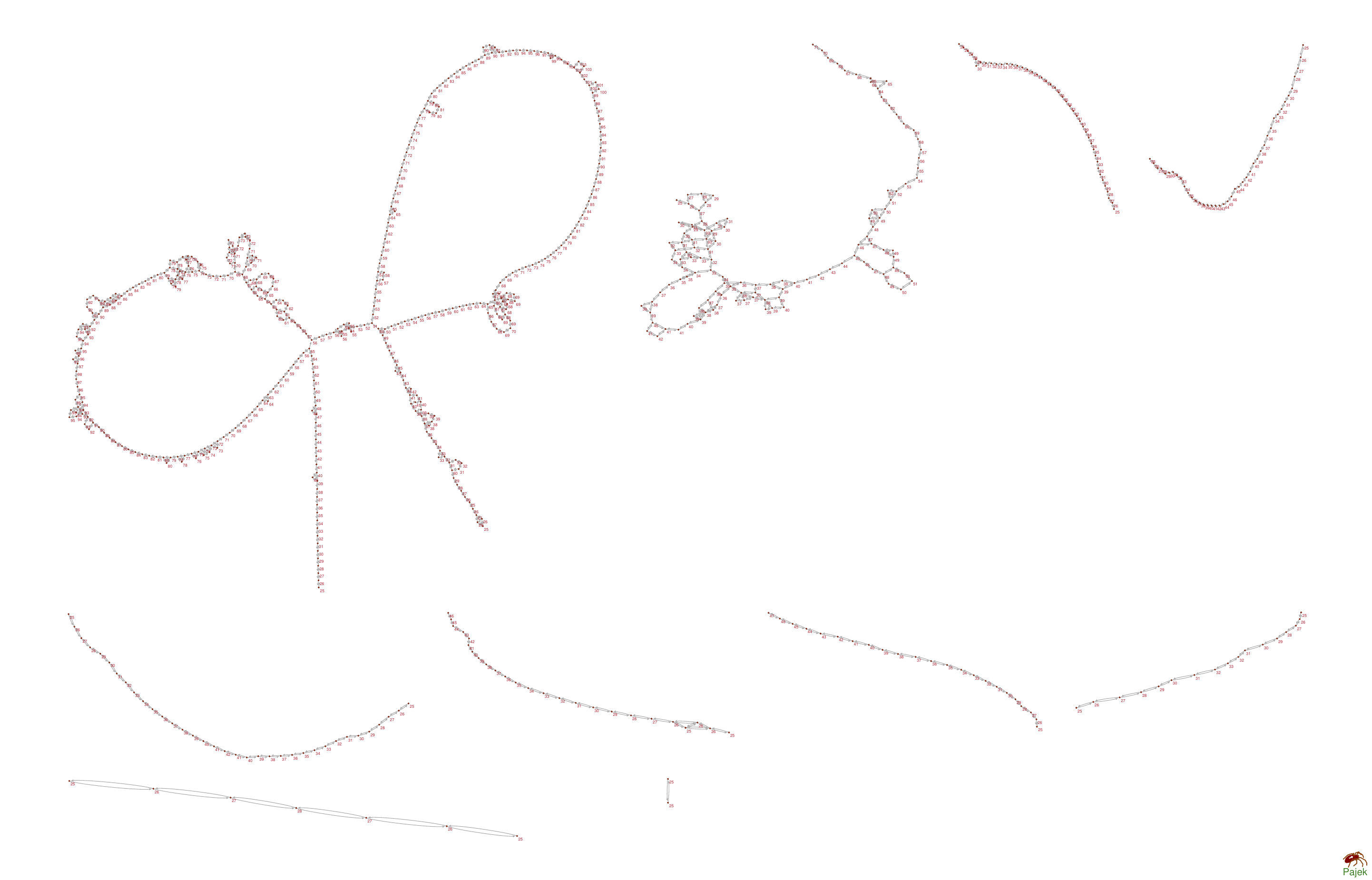}
\end{flushleft}
\caption{Visualization of the remote clusters in the Stanford Web network. The Pajek program package is used \cite{Batagelj1998-ay}. The  vertices are labeled according to their distances from the largest hub in the network. Three of these chains have one of their ends free, and the remaining $7$ chains are parts of long cycles. This Figure is provided as Ancillary file.
}
\label{f10}
\end{figure*}

Figure~\ref{f9} shows an interesting set of the cumulative distributions $P^\text{(cum)}_{\geq m}(s)$ for the undirected projection of a large Stanford Web domain (notice a similar set of cumulative distributions in Fig.~\ref{f6}(b)).   
These empirical cumulative distributions have the $1/s$ asymptotics for $m$ within the range between $10$ and $15$, but for larger $m$, the cumulative distributions become step-like. This step-like shape suggests a specific structure of the remote regions of this network. 
To understand the organization of the connections between the vertices within the remote
regions of the network, we extract the vertices at a distance $m=25$ or beyond from the largest hub in the undirected projection of the giant weakly connected component of the network and edges between them, and visualize the resulting clusters, indicating, for the sake of completeness, the directed edges of the original directed network. In total, there are $714$ vertices in these clusters and $1681$ directed edges. Fig.~\ref{f10} demonstrates this visualization (see also Ancillary file). Notice that almost all these directed edges are reciprocal. Only $3$ directed edges are not reciprocal. This gives the remarkably high fraction $(1681-3)/1681=0.998$ of reciprocal edges in these clusters. The same edge number computation performed including the main part of the network gives a total of $2,234,572$ directed edges, of which $1,649,280$ are not reciprocal, resulting in a much lower value, $0.262$, for the fraction of reciprocal edges.
We see that in this region the network is a set of long chains. Notably, only $3$ of these chains have one of their ends free, and the remaining $7$ chains are parts of long cycles. Loosely speaking, the Web in this remote region is one-dimensional. 


\section{Discussion and conclusions} 
\label{s4}

We have explored one of the basic structural statistical characteristics of the remote regions of complex networks, which previously was known only for uncorrelated networks. We have observed the $s^{-2}$ asymptotics of the distribution $P_{\geq m}(s)$ of the number $s$ of vertices located at distance $m$ or beyond from a randomly chosen vertex in a large set of real and synthetic undirected networks---small worlds---with a surprisingly diverse architectures. Such networks include trees and loopy networks, networks with strong and weak correlations, the one-partite projections of bipartite networks (FP5 net), the undirected projections of directed networks (Stanford Web), collaboration and social networks, the Internet and Web networks.  
This inverse square law is not observed in the networks having no dead ends (vertices of degree $1$) and in finite dimensional networks (power grids, road networks). 

For each of these networks we inspected the product of the cumulative distribution by $N$, $N P^\text{(cum)}_{\geq m}(s)$, which turned out approximately symmetric for all tested cases in the sense that the $x$- and $y$-axes of the plots can be interchanged. 

Moreover, we have revealed that the organization of connections between vertices within the remote regions of networks differs dramatically from the main part of the network, see Fig.~\ref{f10} and Ancillary file. In particular, we have observed a surprisingly high reciprocity $0.998$ of directed edges in the remote region of the Stanford domain of the Web, while the reciprocity equals only $0.262$ in the entire domain.
 
One should emphasize that the theoretical results of Ref.~\cite{dorogovtsev2003metric} for uncorrelated network still do not offer a compelling explanation for the consistent observation of the inverse square law across such a wide spectrum of networks.
The explanation of this law is a challenge for the future work. 
Note that if we assume that the distribution $P_{\geq m}(s)$ has a power-law asymptotics, then, for the divergence of the first moment of this distribution (average number of vertices at distance $m$ or beyond from a randomly chosen vertex), the exponent of this power law must be not greater that $2$. 
Hence the observed exponent $2$ of the asymptotics is the maximum possible value. 

Other challenging directions for the future work are the exploration of remote regions of directed networks and examining the role of the chain structures observed in this work in network processes.


\section*{CRediT authorship contribution statement}

{\bf J.G.~Oliveira:} 
Planning and revision of the manuscript, Designed the study, Carried out numerical simulations, Writing original draft. 
{\bf S.N.~Dorogovtsev:} 
Planning and revision of the manuscript, Designed the study, Carried out analytical and numerical calculations, Writing original draft. 
{\bf J.F.F.~Mendes:} 
Planning and revision of the manuscript.

\section*{Declaration of Competing Interest}

The authors declare that they have no known competing financial interests or personal relationships that could have appeared to influence the work reported in this paper.

\section*{Acknowledgments}

This work was developed within the scope of the project i3N, UIDB/50025/2020 and UIDP/50025/2020, financed by national funds through the FCT/MEC. J.~G.~O. acknowledges fruitful discussions with M.~Argollo~de~Menezes and A.~V\'azquez.












\bibliography{statistics_of_remote}

\begin{thebibliography}{40}
\expandafter\ifx\csname natexlab\endcsname\relax\def\natexlab#1{#1}\fi
\providecommand{\url}[1]{\texttt{#1}}
\providecommand{\href}[2]{#2}
\providecommand{\path}[1]{#1}
\providecommand{\DOIprefix}{doi:}
\providecommand{\ArXivprefix}{arXiv:}
\providecommand{\URLprefix}{URL: }
\providecommand{\Pubmedprefix}{pmid:}
\providecommand{\doi}[1]{\href{http://dx.doi.org/#1}{\path{#1}}}
\providecommand{\Pubmed}[1]{\href{pmid:#1}{\path{#1}}}
\providecommand{\bibinfo}[2]{#2}
\ifx\xfnm\relax \def\xfnm[#1]{\unskip,\space#1}\fi
\bibitem[{Dorogovtsev et~al.(2003)Dorogovtsev, Mendes, and
  Samukhin}]{dorogovtsev2003metric}
\bibinfo{author}{S.~N. Dorogovtsev}, \bibinfo{author}{J.~F.~F. Mendes},
  \bibinfo{author}{A.~N. Samukhin},
\newblock \bibinfo{title}{Metric structure of random networks},
\newblock \bibinfo{journal}{Nucl. Phys. B} \bibinfo{volume}{653}
  (\bibinfo{year}{2003}) \bibinfo{pages}{307}.
\bibitem[{Perra and Fortunato(2008)}]{perra2008spectral}
\bibinfo{author}{N.~Perra}, \bibinfo{author}{S.~Fortunato},
\newblock \bibinfo{title}{Spectral centrality measures in complex networks},
\newblock \bibinfo{journal}{Phys. Rev. E} \bibinfo{volume}{78}
  (\bibinfo{year}{2008}) \bibinfo{pages}{036107}.
\bibitem[{Katz(1953)}]{katz1953new}
\bibinfo{author}{L.~Katz},
\newblock \bibinfo{title}{A new status index derived from sociometric
  analysis},
\newblock \bibinfo{journal}{Psychometrika} \bibinfo{volume}{18}
  (\bibinfo{year}{1953}) \bibinfo{pages}{39}.
\bibitem[{Brin and Page(1998)}]{brin1998anatomy}
\bibinfo{author}{S.~Brin}, \bibinfo{author}{L.~Page},
\newblock \bibinfo{title}{The anatomy of a large-scale hypertextual web search
  engine},
\newblock in: \bibinfo{booktitle}{Proceedings of the Seventh International
  World Wide Web Conference}, \bibinfo{publisher}{Elsevier, Amsterdam},
  \bibinfo{year}{1998}, pp. \bibinfo{pages}{107--117}.
\bibitem[{Page et~al.(1999)Page, Brin, Motwani, and
  Winograd}]{page1999pagerank}
\bibinfo{author}{L.~Page}, \bibinfo{author}{S.~Brin},
  \bibinfo{author}{R.~Motwani}, \bibinfo{author}{T.~Winograd},
  \bibinfo{title}{{The PageRank citation ranking: Bringing order to the web}},
  \bibinfo{type}{Technical Report}, Stanford InfoLab, \bibinfo{year}{1999}.
\bibitem[{Freeman(1977)}]{freeman1977set}
\bibinfo{author}{L.~C. Freeman},
\newblock \bibinfo{title}{A set of measures of centrality based on
  betweenness},
\newblock \bibinfo{journal}{Sociometry} \bibinfo{volume}{40}
  (\bibinfo{year}{1977}) \bibinfo{pages}{35}.
\bibitem[{Newman(2005)}]{newman2005measure}
\bibinfo{author}{M.~E.~J. Newman},
\newblock \bibinfo{title}{A measure of betweenness centrality based on random
  walks},
\newblock \bibinfo{journal}{Social Networks} \bibinfo{volume}{27}
  (\bibinfo{year}{2005}) \bibinfo{pages}{39}.
\bibitem[{Newman(2010)}]{newman2010networks}
\bibinfo{author}{M.~E.~J. Newman}, \bibinfo{title}{Networks: An Introduction},
  \bibinfo{publisher}{Oxford University Press}, \bibinfo{address}{Oxford},
  \bibinfo{year}{2010}.
\bibitem[{Martin et~al.(2014)Martin, Zhang, and
  Newman}]{martin2014localization}
\bibinfo{author}{T.~Martin}, \bibinfo{author}{X.~Zhang},
  \bibinfo{author}{M.~E.~J. Newman},
\newblock \bibinfo{title}{Localization and centrality in networks},
\newblock \bibinfo{journal}{Phys. Rev. E} \bibinfo{volume}{90}
  (\bibinfo{year}{2014}) \bibinfo{pages}{052808}.
\bibitem[{Newman and Girvan(2004)}]{newman2004finding}
\bibinfo{author}{M.~E.~J. Newman}, \bibinfo{author}{M.~Girvan},
\newblock \bibinfo{title}{Finding and evaluating community structure in
  networks},
\newblock \bibinfo{journal}{Phys. Rev. E} \bibinfo{volume}{69}
  (\bibinfo{year}{2004}) \bibinfo{pages}{026113}.
\bibitem[{Estrada and Rodriguez-Velazquez(2005)}]{estrada2005subgraph}
\bibinfo{author}{E.~Estrada}, \bibinfo{author}{J.~A. Rodriguez-Velazquez},
\newblock \bibinfo{title}{Subgraph centrality in complex networks},
\newblock \bibinfo{journal}{Phys. Rev. E} \bibinfo{volume}{71}
  (\bibinfo{year}{2005}) \bibinfo{pages}{056103}.
\bibitem[{Levin and Peres(2017)}]{levin2017markov}
\bibinfo{author}{D.~A. Levin}, \bibinfo{author}{Y.~Peres},
  \bibinfo{title}{Markov Chains and Mixing Times}, volume
  \bibinfo{volume}{107}, \bibinfo{publisher}{American Mathematical Society,
  Providence, RI}, \bibinfo{year}{2017}.
\bibitem[{Samukhin et~al.(2008)Samukhin, Dorogovtsev, and
  Mendes}]{samukhin2008laplacian}
\bibinfo{author}{A.~N. Samukhin}, \bibinfo{author}{S.~N. Dorogovtsev},
  \bibinfo{author}{J.~F.~F. Mendes},
\newblock \bibinfo{title}{{Laplacian spectra of, and random walks on, complex
  networks: Are scale-free architectures really important?}},
\newblock \bibinfo{journal}{Phys. Rev. E} \bibinfo{volume}{77}
  (\bibinfo{year}{2008}) \bibinfo{pages}{036115}.
\bibitem[{Vukadinovi{\'c} et~al.(2002)Vukadinovi{\'c}, Huang, and
  Erlebach}]{vukadinovic2002spectrum}
\bibinfo{author}{D.~Vukadinovi{\'c}}, \bibinfo{author}{P.~Huang},
  \bibinfo{author}{T.~Erlebach},
\newblock \bibinfo{title}{{On the spectrum and structure of Internet topology
  graphs}},
\newblock in: \bibinfo{booktitle}{International Workshop on Innovative Internet
  Community Systems, Lecture Notes in Computer Science}, volume
  \bibinfo{volume}{2346}, \bibinfo{organization}{Springer, Berlin},
  \bibinfo{year}{2002}, pp. \bibinfo{pages}{83--95}.
\bibitem[{Dorogovtsev and Mendes(2022)}]{dorogovtsev2022nature}
\bibinfo{author}{S.~N. Dorogovtsev}, \bibinfo{author}{J.~F.~F. Mendes},
  \bibinfo{title}{The Nature of Complex Networks}, \bibinfo{publisher}{Oxford
  University Press, Oxford}, \bibinfo{year}{2022}.
\bibitem[{Krapivsky and Redner(2001)}]{krapivsky2001organization}
\bibinfo{author}{P.~L. Krapivsky}, \bibinfo{author}{S.~Redner},
\newblock \bibinfo{title}{Organization of growing random networks},
\newblock \bibinfo{journal}{Phys. Rev. E} \bibinfo{volume}{63}
  (\bibinfo{year}{2001}) \bibinfo{pages}{066123}.
\bibitem[{Dorogovtsev and Mendes(2002)}]{dorogovtsev2002evolution}
\bibinfo{author}{S.~N. Dorogovtsev}, \bibinfo{author}{J.~F.~F. Mendes},
\newblock \bibinfo{title}{Evolution of networks},
\newblock \bibinfo{journal}{Adv. Phys.} \bibinfo{volume}{51}
  (\bibinfo{year}{2002}) \bibinfo{pages}{1079}.
\bibitem[{Fronczak et~al.(2004)Fronczak, Fronczak, and
  Ho{\l}yst}]{fronczak2004average}
\bibinfo{author}{A.~Fronczak}, \bibinfo{author}{P.~Fronczak},
  \bibinfo{author}{J.~A. Ho{\l}yst},
\newblock \bibinfo{title}{Average path length in random networks},
\newblock \bibinfo{journal}{Phys. Rev. E} \bibinfo{volume}{70}
  (\bibinfo{year}{2004}) \bibinfo{pages}{056110}.
\bibitem[{van~der Hofstad et~al.(2005)van~der Hofstad, Hooghiemstra, and
  Van~Mieghem}]{van2005distances}
\bibinfo{author}{R.~van~der Hofstad}, \bibinfo{author}{G.~Hooghiemstra},
  \bibinfo{author}{P.~Van~Mieghem},
\newblock \bibinfo{title}{Distances in random graphs with finite variance
  degrees},
\newblock \bibinfo{journal}{Random Struct. Algorithms} \bibinfo{volume}{27}
  (\bibinfo{year}{2005}) \bibinfo{pages}{76}.
\bibitem[{van~der Hofstad et~al.(2007)van~der Hofstad, Hooghiemstra, and
  Znamenski}]{van2007distances}
\bibinfo{author}{R.~van~der Hofstad}, \bibinfo{author}{G.~Hooghiemstra},
  \bibinfo{author}{D.~Znamenski},
\newblock \bibinfo{title}{Distances in random graphs with finite mean and
  infinite variance degrees},
\newblock \bibinfo{journal}{Electron. J. Probab.} \bibinfo{volume}{12}
  (\bibinfo{year}{2007}) \bibinfo{pages}{703}.
\bibitem[{Dorogovtsev et~al.(2006)Dorogovtsev, Mendes, and
  Oliveira}]{dorogovtsev2006degree}
\bibinfo{author}{S.~N. Dorogovtsev}, \bibinfo{author}{J.~F.~F. Mendes},
  \bibinfo{author}{J.~G. Oliveira},
\newblock \bibinfo{title}{Degree-dependent intervertex separation in complex
  networks},
\newblock \bibinfo{journal}{Phys. Rev. E} \bibinfo{volume}{73}
  (\bibinfo{year}{2006}) \bibinfo{pages}{056122}.
\bibitem[{Morohosi(2010)}]{morohosi2010measuring}
\bibinfo{author}{H.~Morohosi},
\newblock \bibinfo{title}{{Measuring the network robustness by Monte Carlo
  estimation of shortest path length distribution}},
\newblock \bibinfo{journal}{Math. Comput. Simul.} \bibinfo{volume}{81}
  (\bibinfo{year}{2010}) \bibinfo{pages}{551}.
\bibitem[{Katzav et~al.(2015)Katzav, Nitzan, Ben-Avraham, Krapivsky, K{\"u}hn,
  Ross, and Biham}]{katzav2015analytical}
\bibinfo{author}{E.~Katzav}, \bibinfo{author}{M.~Nitzan},
  \bibinfo{author}{D.~Ben-Avraham}, \bibinfo{author}{P.~L. Krapivsky},
  \bibinfo{author}{R.~K{\"u}hn}, \bibinfo{author}{N.~Ross},
  \bibinfo{author}{O.~Biham},
\newblock \bibinfo{title}{Analytical results for the distribution of shortest
  path lengths in random networks},
\newblock \bibinfo{journal}{Europhys. Lett.} \bibinfo{volume}{111}
  (\bibinfo{year}{2015}) \bibinfo{pages}{26006}.
\bibitem[{Katzav et~al.(2018)Katzav, Biham, and
  Hartmann}]{katzav2018distribution}
\bibinfo{author}{E.~Katzav}, \bibinfo{author}{O.~Biham}, \bibinfo{author}{A.~K.
  Hartmann},
\newblock \bibinfo{title}{{Distribution of shortest path lengths in subcritical
  Erd{\H{o}}s--R{\'e}nyi networks}},
\newblock \bibinfo{journal}{Phys. Rev. E} \bibinfo{volume}{98}
  (\bibinfo{year}{2018}) \bibinfo{pages}{012301}.
\bibitem[{Tishby et~al.(2022)Tishby, Biham, K{\"u}hn, and
  Katzav}]{tishby2022mean}
\bibinfo{author}{I.~Tishby}, \bibinfo{author}{O.~Biham},
  \bibinfo{author}{R.~K{\"u}hn}, \bibinfo{author}{E.~Katzav},
\newblock \bibinfo{title}{The mean and variance of the distribution of shortest
  path lengths of random regular graphs},
\newblock \bibinfo{journal}{J. Phys. A: Mathematical and Theoretical}
  \bibinfo{volume}{55} (\bibinfo{year}{2022}) \bibinfo{pages}{265005}.
\bibitem[{Dorogovtsev et~al.(2008)Dorogovtsev, Mendes, Samukhin, and
  Zyuzin}]{dorogovtsev2008organization}
\bibinfo{author}{S.~N. Dorogovtsev}, \bibinfo{author}{J.~F.~F. Mendes},
  \bibinfo{author}{A.~N. Samukhin}, \bibinfo{author}{A.~Y. Zyuzin},
\newblock \bibinfo{title}{Organization of modular networks},
\newblock \bibinfo{journal}{Phys. Rev. E} \bibinfo{volume}{78}
  (\bibinfo{year}{2008}) \bibinfo{pages}{056106}.
\bibitem[{Aldous(1990)}]{aldous1990random}
\bibinfo{author}{D.~J. Aldous},
\newblock \bibinfo{title}{The random walk construction of uniform spanning
  trees and uniform labelled trees},
\newblock \bibinfo{journal}{SIAM J. Discret. Math.} \bibinfo{volume}{3}
  (\bibinfo{year}{1990}) \bibinfo{pages}{450}.
\bibitem[{Broder(1989)}]{broder1989generating}
\bibinfo{author}{A.~Z. Broder},
\newblock \bibinfo{title}{Generating random spanning trees},
\newblock in: \bibinfo{booktitle}{30th Annual Symposium on Foundations of
  Computer Science (Research Triangle Park, North Carolina, 30 Oct.---1 Nov.
  1989)}, \bibinfo{publisher}{IEEE, New York}, \bibinfo{year}{1989}, pp.
  \bibinfo{pages}{442--447}.
\bibitem[{Almendral et~al.(2007{\natexlab{a}})Almendral, Oliveira, López,
  Mendes, and Sanjuán}]{almendral2007675}
\bibinfo{author}{J.~A. Almendral}, \bibinfo{author}{J.~G. Oliveira},
  \bibinfo{author}{L.~López}, \bibinfo{author}{J.~F.~F. Mendes},
  \bibinfo{author}{M.~A. Sanjuán},
\newblock \bibinfo{title}{The network of scientific collaborations within the
  {European Framework Programme}},
\newblock \bibinfo{journal}{Physica A: Statistical Mechanics and its
  Applications} \bibinfo{volume}{384} (\bibinfo{year}{2007}{\natexlab{a}})
  \bibinfo{pages}{675--683}.
\bibitem[{Almendral et~al.(2007{\natexlab{b}})Almendral, Oliveira, L\'opez,
  Sanju{\'a}n, and Mendes}]{almendral2007TheIO}
\bibinfo{author}{J.~A. Almendral}, \bibinfo{author}{J.~G. Oliveira},
  \bibinfo{author}{L.~L\'opez}, \bibinfo{author}{M.~A.~F. Sanju{\'a}n},
  \bibinfo{author}{J.~F.~F. Mendes},
\newblock \bibinfo{title}{The interplay of universities and industry through
  the {FP5} network},
\newblock \bibinfo{journal}{New Journal of Physics} \bibinfo{volume}{9}
  (\bibinfo{year}{2007}{\natexlab{b}}) \bibinfo{pages}{183 -- 183}.
\bibitem[{Kunegis(2013)}]{Kunegis2013KONECTTK}
\bibinfo{author}{J.~Kunegis},
\newblock \bibinfo{title}{Konect: the koblenz network collection},
\newblock \bibinfo{journal}{Proceedings of the 22nd International Conference on
  World Wide Web}  (\bibinfo{year}{2013}). \URLprefix
  \url{https://api.semanticscholar.org/CorpusID:15005100}.
\bibitem[{Mislove et~al.(2007)Mislove, Marcon, Gummadi, Druschel, and
  Bhattacharjee}]{mislove-2007-socialnetworks}
\bibinfo{author}{A.~Mislove}, \bibinfo{author}{M.~Marcon},
  \bibinfo{author}{K.~P. Gummadi}, \bibinfo{author}{P.~Druschel},
  \bibinfo{author}{B.~Bhattacharjee},
\newblock \bibinfo{title}{{Measurement and Analysis of Online Social
  Networks}},
\newblock in: \bibinfo{booktitle}{Proceedings of the 5th ACM/Usenix Internet
  Measurement Conference (IMC'07)}, \bibinfo{address}{San Diego, CA},
  \bibinfo{year}{2007}.
\bibitem[{Viswanath et~al.(2009)Viswanath, Mislove, Cha, and
  Gummadi}]{viswanath-2009-activity}
\bibinfo{author}{B.~Viswanath}, \bibinfo{author}{A.~Mislove},
  \bibinfo{author}{M.~Cha}, \bibinfo{author}{K.~P. Gummadi},
\newblock \bibinfo{title}{On the evolution of user interaction in facebook},
\newblock in: \bibinfo{booktitle}{Proceedings of the 2nd ACM SIGCOMM Workshop
  on Social Networks (WOSN'09)}, \bibinfo{year}{2009}.
\bibitem[{cai(2013)}]{caida_router}
\bibinfo{title}{{CAIDA Skitter Router-Level Topology and Degree Distribution,
  from April 21 to May 8, 2003}},
  \bibinfo{howpublished}{\url{https://www.caida.org/catalog/datasets/router-adjacencies/}},
  \bibinfo{year}{2013}.
\bibitem[{as_(2013)}]{as_caida}
\bibinfo{title}{The dataset contains 122 {CAIDA AS} graphs, from {January} 2004
  to {November} 2007},
  \bibinfo{howpublished}{\url{http://www.caida.org/data/active/as-relationships/}},
  \bibinfo{year}{2013}.
\bibitem[{Pastor-Satorras and Vespignani(2004)}]{pastor2004evolution}
\bibinfo{author}{R.~Pastor-Satorras}, \bibinfo{author}{A.~Vespignani},
  \bibinfo{title}{Evolution and Structure of the Internet: A Statistical
  Physics Approach}, \bibinfo{publisher}{Cambridge University Press,
  Cambridge}, \bibinfo{year}{2004}.
\bibitem[{Watts and Strogatz(1998)}]{Watts1998-ve}
\bibinfo{author}{D.~J. Watts}, \bibinfo{author}{S.~H. Strogatz},
\newblock \bibinfo{title}{Collective dynamics of 'small-world' networks},
\newblock \bibinfo{journal}{Nature} \bibinfo{volume}{393}
  (\bibinfo{year}{1998}) \bibinfo{pages}{440--442}.
\bibitem[{Leskovec et~al.(2009)Leskovec, Lang, Dasgupta, and
  Mahoney}]{Leskovec2009-vs}
\bibinfo{author}{J.~Leskovec}, \bibinfo{author}{K.~J. Lang},
  \bibinfo{author}{A.~Dasgupta}, \bibinfo{author}{M.~W. Mahoney},
\newblock \bibinfo{title}{Community structure in large networks: Natural
  cluster sizes and the absence of large well-defined clusters},
\newblock \bibinfo{journal}{Internet Math.} \bibinfo{volume}{6}
  (\bibinfo{year}{2009}) \bibinfo{pages}{29--123}.
\bibitem[{Palla et~al.(2007)Palla, Farkas, Pollner, Der{\'e}nyi, and
  Vicsek}]{palla2007directed}
\bibinfo{author}{G.~Palla}, \bibinfo{author}{I.~J. Farkas},
  \bibinfo{author}{P.~Pollner}, \bibinfo{author}{I.~Der{\'e}nyi},
  \bibinfo{author}{T.~Vicsek},
\newblock \bibinfo{title}{Directed network modules},
\newblock \bibinfo{journal}{New J. Phys.} \bibinfo{volume}{9}
  (\bibinfo{year}{2007}) \bibinfo{pages}{186}.
\bibitem[{Batagelj(1998)}]{Batagelj1998-ay}
\bibinfo{author}{V.~Batagelj},
\newblock \bibinfo{title}{Mrvar: Pajek - program for large network analysis},
\newblock \bibinfo{journal}{Connections} \bibinfo{volume}{21}
  (\bibinfo{year}{1998}) \bibinfo{pages}{47--57}.

\end{thebibliography}

\end{document}